\documentclass[sigconf, screen=true, usenames, 10pt]{acmart}
\pdfoutput=1
\usepackage{mdframed}
\usepackage{bbm}
\usepackage{graphicx}   
\usepackage{hyperref}   

\usepackage{amsmath,amssymb}    
\usepackage{thm-restate}
\usepackage{mdwlist}
\usepackage{mathtools}
\usepackage{xspace}
\usepackage{verbatim}   
\usepackage{xcolor}
\usepackage{makecell}
\usepackage[mathscr]{euscript}
\usepackage{tikz}
\usepackage{relsize}
\usepackage{booktabs}	
\usepackage{arydshln}
\usepackage{tikz-qtree}
\usepackage{caption}
\usepackage{subcaption}
\usepackage{multirow}
\usetikzlibrary{arrows.meta}
\usetikzlibrary{fit,shapes,trees,shapes.geometric}
\usetikzlibrary{patterns,decorations.pathreplacing,calc}
\usetikzlibrary{matrix, positioning, arrows}
\usetikzlibrary{chains,shapes.multipart}
\usetikzlibrary{shapes,calc}
\usetikzlibrary{automata}
\usepackage[linesnumbered,algoruled, lined, noend]{algorithm2e}

\DeclareRobustCommand{\newtt}{\fontfamily{lmtt}\selectfont}
\DeclareRobustCommand\oldtt[1]{{\mbox{\bfseries \newtt#1}}}

\usepackage[normalem]{ulem}
\usepackage{listings,lstautogobble}

\DeclareMathAlphabet\mathbfcal{OMS}{cmsy}{b}{n}

\newcommand{\cut}[1]{}   
\makeatletter
\newcommand{\mytag}[2]{%
	\text{#1}%
	\@bsphack
	\protected@write\@auxout{}%
	{\string\newlabel{#2}{{#1}{\thepage}}}%
	\@esphack
}
\makeatother

\newcommand{\introparagraph}[1]{\noindent {\bf \em #1.}}  
\newcommand{\noemintroparagraph}[1]{\noindent {\bf  #1.}}

\newcommand{\setof}[2]{\{{#1}\mid{#2}\}}        

\usepackage{aliascnt} 

\newtheorem{proposition}{Proposition}

\newtheorem{lemma}{Lemma}
\newtheorem{example}{Example}

\providecommand{\mA}[0]{\mathbfcal{A}}

\providecommand{\mL}[0]{\mathcal{L}}

\providecommand{\mB}[0]{\mathbfcal{B}}

\providecommand{\sN}[0]{{N}}
\providecommand{\sNR}[0]{{N_R}}

\providecommand{\sNS}[0]{{N_S}}
\providecommand{\mtwopath}[0]{\ddot{Q}}

\providecommand{\tOUT}[0]{\texttt{OUT}}
\providecommand{\bD}[0]{\mathbf{D}}

\providecommand{\domain}[0]{\mathbf{dom}}

\newcommand{\hlrone}[1]{{\color{black} {#1}}}
\newcommand{\hlrtwo}[1]{{\color{black} {#1}}}

\newcommand{\eat}[1]{}

\newcommand{\cC}[0]{C}

\newcommand{\cab}[4]{  {\small{\color{#2} #1}, {\color{#4} #3}}}

\hypersetup{
	colorlinks=true,
	citecolor=red,
	linkcolor=red}

\usepackage{libertine}
\usepackage[utf8]{inputenc}
\usepackage[T1]{fontenc}
\usepackage{microtype}
\usepackage{enumitem}

\begin{document}

	\title{Fast Join Project Query Evaluation using Matrix Multiplication}
	\author{Shaleen Deep}
	\affiliation{\institution{University of Wisconsin-Madison}\country{United States of America}}
    
	\email{shaleen@cs.wisc.edu}
	\author{Xiao Hu}
	\affiliation{\institution{Duke University}\country{United States of America}}
	\email{xh102@duke.edu.com}
	\author{Paraschos Koutris}
	\affiliation{\institution{University of Wisconsin-Madison}\country{United States of America}}
	\email{paris@cs.wisc.edu}
	
	\begin{abstract}
	In the last few years, much effort has been devoted to developing join algorithms in order to achieve worst-case optimality for join queries over relational databases. Towards this end, the database community has had considerable success in developing succinct algorithms that achieve worst-case optimal runtime for full join queries, i.e the join is over all variables present in the input database.
	However, not much is known about join evaluation with {\em projections} beyond some simple techniques of pushing down the projection operator in the query execution plan. Such queries have a large number of applications in entity matching, graph analytics and searching over compressed graphs. In this paper, we study how a class of join queries with projections can be evaluated faster using worst-case optimal algorithms together with matrix multiplication. 
Crucially, our algorithms are parameterized by the output size of the final result, allowing for choice of the best execution strategy. We implement our algorithms as a subroutine and compare the performance with state-of-the-art techniques to show they can be improved upon by as much as 50x. More importantly, our experiments indicate that matrix multiplication is a useful operation that can help speed up join processing owing to highly optimized open source libraries that are also highly parallelizable.
\end{abstract}

	\maketitle
\section{Introduction} \label{sec:intro}

In this paper, we study the problem of evaluating join queries where the join result does not contain all the variables in body of the query. In other words, some of the variables have been {\em projected out} of the join result. The simplest way to evaluate such a query is to first compute the full join, and then make a linear pass over the result, project each tuple and remove the duplicates. While this approach is conceptually simple, it relies on efficient {\em worst-case optimal join algorithms} for full queries, which have recently been developed in a series of papers~\cite{AGM,skewstrikesback, ngo2012worst, veldhuizenleapfrog}.
The main result in this line of work is a class of algorithms that run in time $O({| \bD |}^{\rho^*} + |\tOUT|)$, where $\bD$ is the database instance and $\rho^*$ is the optimal fractional edge cover of the query~\cite{AGM}. In the worst case, there exists a database $\bD$ such that $|\tOUT| = {|\bD|}^{\rho^*}$. In practice, most query optimizers create a query plan by pushing down projections in the join tree.

\begin{example} \label{ex:1}
\hlrone{Consider relation $R(x,y)$ of size $\sN$ that represents a social network graph where an edge between two users $x$ and $y$ denotes that $x$ and $y$ are friends. We wish to enumerate all users pairs who have at least one friend in common ~\cite{misra2012community}. This task is equivalent to the query $\mtwopath(x,z) = R(x,y), R(z,y)$, which corresponds to the following \texttt{SQL} query: $\oldtt{SELECT DISTINCT R1.x}$ $\oldtt{, R2.x}$ $\oldtt{FROM R1 as R, R2 as R}$ $\oldtt{WHERE R1.y = R2.y}$. \\
Suppose that the graph contains a small (constant) number of communities and the users are spread evenly across them. Each community has $O(\sqrt{\sN})$ users, and there exists an edge between most user pairs within the same community. In this case, the full join result is $\Theta(\sN^{3/2})$ but $|\mtwopath(\bD)| = \Theta(\sN)$}.
\end{example}

As the above example demonstrates, using worst-case optimal join algorithms can lead to an intermediate output that can be much larger than the final result after projection, especially if there are many duplicate tuples. Thus, we ask whether it is possible to design faster algorithms that can skip the construction of the full result when this is large and as a result speed up the evaluation. Ideally, we would like to have algorithms that run faster than worst-case optimal join algorithms, are sensitive to the output of projected result, and do not require large main memory during execution. 

In this paper, we show how to achieve the above goal for a fundamental class of join queries called {\em star joins}. Star joins are join queries where every relation is joined on the same variable. The motivation to build faster algorithms for star joins with projection is not limited to faster query execution in DBMS systems. We present next a list of three applications that benefit from these faster algorithms.

\smallskip
\introparagraph{Set Similarity} Set similarity is a fundamental operation in many applications such as entity matching and recommender systems. Here, the goal is to return all pairs of sets such that have contain at least $c$ common elements. Recent work~\cite{deng2018overlap} gave the first output-sensitive algorithm that enumerates all similar sets in time $O(|\bD|^{2-\frac{1}{c}} \cdot |\tOUT|^{\frac{1}{2c}})$.  As the value of $c$ increases, the running time tends to $O(|\bD|^2)$. The algorithm also requires $O(|\bD|^{2-\frac{1}{c}} \cdot |\tOUT|^{\frac{1}{2c}})$ space.  We improve the running time and the space requirement of the algorithm for a large set of values that $|\tOUT|$ can take, for all $c$.

\smallskip
\introparagraph{Set Containment} Efficient computation of set containment joins over set-value attributes has been extensively studied in the literature. A long line of research~\cite{jampani2005using, yang2018efficient, kunkel2016piejoin,luo2015efficient} has developed a trie-based join method where the algorithm performs an efficient blocking step that prunes away most of the set verifications. However, the verification step is a simple set merging based method that checks if set $\hlrone{t \subseteq u}$, which can be expensive. We show that for certain datasets, our algorithm can identify set containment relationships much faster than state-of-the-art techniques.

\smallskip
\introparagraph{Graph Analytics} In the context of graph analytics, the graph to be analyzed is often defined as a {\em declarative} query over a relational schema~\cite{graphgen2017, graphgen2015, graphgen2017adaptive, spartex}. For instance, consider the DBLP dataset, which stores which authors write which papers through a table $R(author, paper)$. To analyze the relationships between co-authors, we can extract the {\em co-author graph}, which we can express as the view $V(x,y) = R(x,p), R(y,p)$. Recent work~\cite{graphgen2017} has proposed compression techniques where a preprocessing step generates a succinct representation of $V(x,y)$. However, these techniques require a very expensive pre-processing step, rely on heuristics, and do not provide any formal guarantees on the running time. In the context of querying data through \textsf{APIs}, suppose that we want to support an API where a user checks whether authors $a_1$ and $a_2$ have co-authored a paper. This is an example of a {\em boolean query}. In this scenario, the view $R(x,p), R(y,p)$ is implicit and not materialized. Since such an API may handle thousands of requests per second, it is beneficial to batch $B$ queries together, and evaluate them at once. We show that our algorithms can lead to improved performance by minimizing user latency and resource usage. 

\smallskip
\introparagraph{Our contribution} In this paper, we show how to evaluate star join queries with projection using output-sensitive algorithms. We summarize our technical contribution below.

\begin{enumerate}
	
	\item Our main contribution (\autoref{sec:joinproject}) is an output-sensitive algorithm that evaluates star join queries with projection . We use worst-case optimal joins and matrix multiplication as two fundamental building blocks to split the join into multiple subjoin queries which are evaluated separately. This technique was initially introduced in~\cite{amossen2009faster}, but their runtime analysis is incorrect for certain regimes of the output size. We improve and generalize the results via a more careful application of (fast) matrix multiplication.
	
	\item We show (\autoref{sec:application}) how to exploit the join query algorithms for the problems of set similarity, set containment, join processing and boolean set intersection.  Our algorithms also improve the best known preprocessing time bounds for creating offline data structures for set intersection problems~\cite{deep2017compressed} and compressing large graphs~\cite{graphgen2017}. In addition, we can show that our approach is much more amenable to parallelization.
	
	\item We develop (\autoref{sec:optimizer} and \autoref{sec:implementation}) a series of optimization techniques that address the practical challenges of incorporating matrix multiplication algorithms into join processing.
	
	\item We implement our solution as an in-memory prototype and perform a comprehensive benchmarking to demonstrate the usefulness of our approach (\autoref{sec:exp}). We show that our algorithms can be used to improve the running time for set similarity, set containment, join processing and boolean query answering over various datasets for both single threaded and multithreaded settings. Our experiments indicate that matrix multiplication can achieve an order of magnitude speedup on average and upto $50\times$ speedup over the best known baselines.
	
\end{enumerate}


\section{Problem Setting}
\label{sec:framework}

In this section we present the basic notions and terminology, and then define the problems we study in this paper.

\subsection{Problem Definitions}

In this paper, we will focus on the 2-path query, which consists of a binary join followed by a projection:
\[ \mtwopath(x,z) = R(x,y), S(z, y)\]
and its generalization as a {\em star join}:
\begin{equation*} \label{eq:star}
Q^{\star}_k(x_1, x_2, \dots, x_k) = R_1(x_1, y), R_2(x_2, y), \ldots, R_k(x_k, y). 
\end{equation*}

We will often use the notation $\domain(x)$ to denote the constants that the values that variable $x$ can take. We use $Q(\bD)$ to denote the result of the query $Q$ over input database $\bD$, or also or $\tOUT$ when it is clear from the context.

Apart from the above queries, the following closely related problems will also be of interest.

\smallskip
\noindent \noemintroparagraph{Set Similarity (SSJ)} 
In this problem, we are given two families of sets represented by the binary relations $R(x,y)$ and $\hlrtwo{S(z,y)}$.
Here, $R(x,y)$ means set $x$ contains element $y$, and $\hlrtwo{S(z,y)}$ means set $z$ contains element $y$.
Given an integer $c \geq 1$, the set similarity join is defined as
$$ \setof{(a,b)}{ |\pi_y(\sigma_{x=a} (R)) \cap \pi_y(\sigma_{z=b}(S))| \geq c} $$
In other words, we want to output the pairs of sets where their intersection size is at least $c$. When $c=1$, SSJ becomes equivalent to the 2-path query $\mtwopath$. The generalization of set similarity to more than two relations can be defined in a similar fashion. 
Previous work~\cite{deng2018overlap} only considered the unordered version of SSJ. The {\em ordered} version simply enumerates $\tOUT$ in  decreasing order of similarity. This allows users to see the most similar pairs first instead of enumerating output tuples in arbitrary order. 

\smallskip
\noindent \noemintroparagraph{Set Containment (SCJ)} Similar to SSJ, given two families of sets represented by the relations $R,S$, we want to output
$$ \setof{(a,b)}{ \pi_y(\sigma_{x=a} (R)) \subseteq \pi_y(\sigma_{z=b}(S)) } $$
In other words, we want to output the pairs of sets where one set is contained in the other.

\smallskip
\noindent \noemintroparagraph{Boolean Set Intersection (BSI)} In this problem, we are given again two families of sets represented by the relations $R,S$. Then, for every input pair two sets $a,b$, we want to answer the following boolean CQ which asks whether the two sets have a non-empty intersection: $ Q_{ab}() = R(a,y), \hlrtwo{S(b,y)}$.
If we also want to output the actual intersection, we can use the slightly modified CQ $ \bar{Q}_{ab}(y) = R(a,y), \hlrtwo{S(b,y)}$, which does not project the join variable.
The boolean set intersection problem has been a subject of great interest in the theory community~\cite{cohen2010hardness, patrascu2010distance, afshani2016data, Cohen2010, deep2017compressed} given its tight connections with distance oracles and reachability problems in graphs.

\smallskip
In order to study the complexity of our algorithms, we will use the uniform-cost RAM 
model~\cite{hopcroft1975design}, where data values as well as pointers to
databases are of constant size. Throughout the paper, all complexity results are 
with respect to data complexity where the query is assumed fixed.

\subsection{Matrix Multiplication}

Let $A$ be a $U \times V$ matrix and $C$ be a $V \times W$ matrix over a field $\mathcal{F}$. 
We use $A_{i,j}$ to denote the entry of $A$ located in row $i$ and column $j$. The matrix product $AC$ is a $U \times W$ matrix with entries $(AC)_{i,j} = \sum_{k=1}^{V} A_{i,k} C_{k,j}$. 

\smallskip
\introparagraph{Join-Project as Matrix Multiplication} It will be convenient to view the 2-path query as a matrix computation.
Let $\mA, \mB$ be the adjacency matrices for relations $R,S$ respectively: this means that $\mA_{i,j} = 1$ if and only if tuple $(i, j) \in R$ (similarly for $S$).  Observe that although each relation has size at most $|\bD|$, the input adjacency matrix can be as large as $|\bD|^2$.
The join output result $\mtwopath(D)$ can now be expressed as the matrix product $\mA \cdot \mB$, where matrix multiplication is performed over the boolean field.

%
%

\introparagraph{Complexity}
Multiplying two square matrices of size $n$ trivially takes time $O(n^3)$, but a long line of research on fast matrix multiplication has dropped the complexity to $O(n^\omega)$, where $2 \leq \omega< 3$. The current best known value is $\omega = 2.373$~\cite{gall2018improved}, but it is believed that the actual value is $2+o(1)$.

We will frequently use the following folklore lemma; we have added its proof for completeness.

\begin{lemma} \label{lem:matrix:multiplication}
Let $\omega$ be any constant such that we can multiply two $n \times n$ matrices in time $O(n^{\omega})$.  Then, two matrices of size $U \times V$ and $V \times W$ can be multiplied in time $M(U,V,W) = O(UVW \beta^{\omega-3})$, where $\beta = \min \{U,V,W\}$. 
\end{lemma}
\begin{proof}
	\hlrone{Assume w.l.o.g. that $\beta$ divides $\alpha = UVW/\beta$. Since $\beta$ is the smallest dimension we can divide the matrices into $\alpha/\beta^2$ submatrices of size $ \beta \times \beta$, which can be multiplied using $O(\beta^\omega)$ operations.}
\end{proof}

For the theoretically best possible $\omega = 2+o(1)$, rectangular fast matrix multiplication can be done in time $O(UVW/\beta)$.

\subsection{Known Results} \label{sec:known}

Ideally we would like to compute $Q^\star_k$ in time linear to the size of the input and output. However, \cite{DBLP:conf/csl/BaganDG07} showed that $\mtwopath$ cannot be evaluated in time $O(|\tOUT|)$ assuming that exponent $\omega$ in matrix multiplication is greater than two. 

A straightforward way to compute any query that is a join followed by a projection is to compute the join using any worst-case optimal algorithm, and then deduplicate to find the projection. This gives the following baseline result.

\begin{proposition}[~\cite{skewstrikesback,ngo2012worst}]\label{prop:basic}
Any CQ $Q$ with optimal fractional edge cover $\rho^*$ can be computed in time $O(|\bD|^{\rho^*})$.
\end{proposition}

\autoref{prop:basic} implies that we can compute the the star query $Q^\star_k$ in time $O(|\bD|^k)$, where $k$ is the number of joins. However, the algorithm is oblivious of the actual output $\tOUT$ and will have the same worst-case running time even if $\tOUT$ is much smaller than $|\bD|^k$ -- as it happens often in practice.
To circumvent this issue, \cite{amossen2009faster} showed the following output sensitive bound that uses only combinatorial techniques:

\begin{lemma}[~\cite{amossen2009faster}] \label{lem:star:basic}
$Q^\star_k$ can be computed in time $O(|\bD| \cdot |\tOUT|^{1 - \frac{1}{k}})$.
\end{lemma}
For $k=2$, the authors make use of fast matrix multiplication to improve the running time to $\tilde{O}(N^{0.862} \cdot |\tOUT|^{0.408} + |\bD|^{2/3} \cdot |\tOUT|^{2/3})$. In the next section, we will discuss the flaws in the proof of this result in detail.

\eat{Observe that the claimed result implies a sub-linear time algorithm if $|\tOUT| = o(|\bD|^{1/2})$. We now point out two errors in their analysis rendering the claimed complexity result incorrect:

\begin{enumerate}
\item  The algorithm uses matrix multiplication to compute the product between two matrices of size $x \times y$ and $y \times z$ in time $\tilde{O}(C(x,y,z))$ where $C(x,y,z)$ denotes the minimum number of arithmetic operations needed in order to multiply the two matrices. However, $C(x,y,z)$ measures only the number of {\em multiplications} performed, and implicitly assumes that the matrices are already computed. Thus, the analysis fails to take into account the time to construct the matrices and converting them into an adjacency list form as required for the best known formula for $C(x,y,z)$~\cite{huang1998fast} to be applicable.
	
\item The algorithm identifies suitable values for degree threshold $\Delta_b$ and $\Delta_{ac}$. However, the value of $\Delta_b$ that minimizes the running time happen to be smaller than one. This cannot be the case given the constraint that degree thresholds need to be at least one and at most $|\bD|$.
\end{enumerate}}

\section{Computing Join-Project} 
\label{sec:joinproject}

In this section, we describe our main technique and its theoretical analysis. 

\subsection{The 2-Path Query} \label{sec:joinproject:twopath}
Consider the query $\mtwopath(x,z) = R(x,y), S(z,y)$. Let $\sNR$ and $\sNS$ denote the cardinality of relations $R$ and $S$ respectively. Without loss of generality, assume that $\sNS \leq \sNR$. For now, assume that we know the output size $|\tOUT|$; we will show how to drop this assumption later.

We will also assume that we have removed any tuples that do not contribute to the query result, which we can do during a linear time preprocessing step.

\introparagraph{Algorithm}
\hlrone{Our algorithm follows the idea of partitioning the input tuples based on their degree as introduced in~\cite{amossen2009faster}, but it differs on the choice of threshold parameters}. It is parametrized by two integer constants $\Delta_1, \Delta_2 \geq 1$.
It first partitions each relation into two parts, $R^-, R^+$ and $S^-, S^+$:
\begin{align*}
R^- & = \{R(a,b) \mid |\sigma_{x=a} R(x,y)| \leq \Delta_2 \text{ or } |\sigma_{y=b} S(z, y)| \leq \Delta_1 \} \\  
S^- & = \{S(c,b) \mid |\sigma_{z=c} S(z,y)| \leq \Delta_2 \text{ or } |\sigma_{y=b} S(z,y)| \leq \Delta_1 \}
\end{align*}

In other words, $R^-, S^-$ include the tuples that contain at least one value with low degree. 
$R^+, S^+$ contain the remaining tuples from $R,S$ respectively. 
\autoref{algo:two:path} describes the detailed steps for computing the join. It proceeds by performing a (disjoint) union of the following results:
\begin{enumerate}
	\item \label{light} Compute $R^- \Join S$ and $R \Join S^-$ using any worst-case optimal join algorithm, then project. 
	\item Materialize $R^+, S^+$ as two rectangular matrices and use matrix multiplication to compute their product. 
\end{enumerate}

Intuitively, the "light" values are handled by standard join techniques, since they will not result in a large intermediate result before the projection. On the other hand, since the "heavy" values will cause a large output, it is better to compute their result directly  using (fast) matrix multiplication.

\begin{example}
	\hlrtwo{Consider relation $R$ and $S$ as shown below.} 
	
	\hspace{3em}
	\begin{minipage}{\linewidth}
		\scalebox{.85}{\begin{tikzpicture}
			
			\begin{scope}[fill opacity=1]
			
			\node at (-1,1) {Relation $R$};							
			\node at (-2,0.5) {\color{brown}$\mathbf{x}$};		
			\node at (0,0.5) {\color{orange}$\mathbf{y}$};		
			\node (x1) at (-2,0) {$1$};
			\node (x2) at (-2,-0.5) {$2$};
			\node (x3) at (-2,-1) {$3$};
			\node (x4) at (-2,-1.5) {$4$};
			\node (x5) at (-2,-2) {$5$};
			\node (x6) at (-2,-2.5) {$6$};

			\node (y1) at (0,0) {$1$};
			\node (y2) at (0,-0.5) {$2$};
			\node (y3) at (0,-1) {$3$};
			\node (y4) at (0,-1.5) {$4$};
			\node (y5) at (0,-2) {$5$};
			\node (y6) at (0,-2.5) {$6$};			
			
			\draw[-, color=red] (x1) to (y6);
			\draw[-, color=red] (x2) to (y1);
			\draw[-, color=red] (x2) to (y2);
			\draw[-, color=red] (x3) to (y5);
			\draw[-, color=red] (x3) to (y3);
			\draw[-] (x4) to (y4);
			\draw[-, color=red] (x4) to (y1);
			\draw[-] (x4) to (y6);
			\draw[-] (x5) to (y4);
			\draw[-] (x5) to (y5);
			\draw[-] (x5) to (y6);			
			\draw[-] (x6) to (y4);
			\draw[-] (x6) to (y5);
			\draw[-, color=red] (x6) to (y2);						
			
			\node at (3,1) {Relation $S$};										
			\node at (2,0.5) {\color{cyan} $\mathbf{z}$};		
			\node at (4,0.5) {\color{orange}$\mathbf{y}$};		
			\node (z1) at (2,0) {$1$};
			\node (z2) at (2,-0.5) {$2$};
			\node (z3) at (2,-1) {$3$};
			\node (z4) at (2,-1.5) {$4$};
			\node (z5) at (2,-2) {$5$};
			\node (z6) at (2,-2.5) {$6$};

			\node (yy1) at (4,0) {$1$};
			\node (yy2) at (4,-0.5) {$2$};
			\node (yy3) at (4,-1) {$3$};
			\node (yy4) at (4,-1.5) {$4$};
			\node (yy5) at (4,-2) {$5$};
			\node (yy6) at (4,-2.5) {$6$};			
			
			\draw[-, color=green] (z1) to (yy6);
			\draw[-, color=green] (z1) to (yy2);
			\draw[-, color=green] (z2) to (yy6);
			\draw[-, color=green] (z2) to (yy3);
			\draw[-, color=green] (z3) to (yy3);
			\draw[-] (z4) to (yy4);
			\draw[-] (z4) to (yy5);
			\draw[-, color=green] (z4) to (yy1);
			\draw[-] (z5) to (yy4);
			\draw[-] (z5) to (yy5);
			\draw[-] (z5) to (yy6);			
			\draw[-, color=green] (z6) to (yy2);
			\draw[-] (z6) to (yy5);
			\draw[-] (z6) to (yy6);

			\end{scope}	
			\end{tikzpicture}
		}
	\end{minipage}
	
	\hlrtwo{Suppose $\Delta_{1} = \Delta_{2} = 2$. Then, all the edges marked in red (green) form relation $R^-(S^-)$. $R^- \Join S$ and $R \Join S^-$ can now be evaluated using any worst-case optimal algorithm. The remaining edges consist of values that are heavy. Thus, we construct matrices $M_1$ and $M_2$ encoding all heavy tuples. The resulting matrix product $M$ shows all the heavy output tuples with their corresponding counts.}
	
	\begin{minipage}{\linewidth}
		\begin{tikzpicture}
	\matrix (K) [matrix of nodes,row sep=-\pgflinewidth, nodes={draw, fill=blue!10}]
	{
		& |[fill=orange!40]|4 & |[fill=orange!40]|5 & |[fill=orange!40]|6 \\
		|[fill=brown!40]|4 & 1 & 0 & 1 \\
		|[fill=brown!40]|5 & 1 & 1 & 1 \\
		|[fill=brown!40]|6 & 1 & 1 & 0 \\
	};
	
	\node[below= 1.5em of K-3-2.south] (l1) {\qquad matrix $M_1$};
	
	\node[right= -0.5em of K-3-3.east] (l2) {\qquad $\times$};

	\matrix (L) [right=1.5em of K,matrix of nodes,row sep=0em, nodes={draw, fill=blue!10}]
	{
		& |[fill=cyan!40]|4 & |[fill=cyan!40]|5 & |[fill=cyan!40]|6 \\
		|[fill=orange!40]|4 & 1 & 1 & 0 \\
		|[fill=orange!40]|5 & 1 & 1 & 1 \\
		|[fill=orange!40]|6 & 0 & 1 & 1 \\
	};
	
	\node[below= 1.5em of L-3-2.south] (l3) {\qquad matrix $M_2$};
	
	\node[right= -0.5em of L-3-3.east] (l4) {\qquad $=$};
	
	\matrix (M) [right=1.5em of L,matrix of nodes,row sep=0em, nodes={draw, fill=blue!10}]
	{
		& |[fill=cyan!40]|4 & |[fill=cyan!40]|5 & |[fill=cyan!40]|6 \\
		|[fill=brown!40]|4 & 1 & 2 & 1 \\
		|[fill=brown!40]|5 & 2 & 3 & 2 \\
		|[fill=brown!40]|6 & 2 & 2 & 3 \\
	};
	\node[below= 1.5em of M-3-2.south] (l5) {\qquad matrix $M$};
	\end{tikzpicture} 

		\end{minipage}

\end{example}

\begin{algorithm} [!t]
	\SetCommentSty{textsf}
	\DontPrintSemicolon 	
	\BlankLine
	$R^{-} \leftarrow \{R(a,b) \mid |\sigma_{x=a} R(x,y)| \leq \Delta_2 \text{ or } |\sigma_{y=b} R(x, y)| \leq \Delta_1\}$, $R^+ \leftarrow R \setminus R^-$ \;
	$S^{-} \leftarrow \{S(c,b) \mid |\sigma_{z=c} S(z,y)| \leq \Delta_2 \text{ or } |\sigma_{y=b} S(z,y)| \leq \Delta_1\}$, $S^+ \leftarrow S \setminus S^-$ \;
	$T \leftarrow (R^- \Join S) \cup (R \Join S^-)$ \label{line:first} \tcc*{use wcoj}
	$M_1(x,y) \leftarrow R^+ \text{ adj matrix}, M_2(y,z) \leftarrow S^+ \text{ adj matrix}$ \;
	$M \leftarrow M_1 \times M_2$ \tcc*{matrix multiplication}
	$T \leftarrow T \cup \{ (a,c) \mid M_{ac} > 0 \}$ \;
	\KwRet{$T$}
	\caption{Computing $\pi_{xz} R(x,y) \Join S(z,y)$}
	\label{algo:two:path}
\end{algorithm}

\hlrtwo{\introparagraph{Correctness} Consider an output tuple $(a,c)$. If there exists no $b$ such that $(a,b) \in R$ and $(c, b) \in S$, then such a pair cannot occur in the output since it will not occur in $R^- \Join S, R \Join S^-$ or $M$. Now suppose that $(a,c)$ has at least one witness $b$ such that $(a,b) \in R$ and $(c,b) \in S$. If $b$ is light in relation $R$ or $S$, then at least one of $(a,b)$ or $(c,b)$ will be included in $R^-$ or $S^-$ and the output tuples will be discovered in the join of $R^- \Join S$ or $R \Join S^-$. Similarly, if the degree of $a$ or $c$ is at most $\Delta_{2}$ in relation $R$ or $S$ respectively, the output tuple will be found in $R^- \Join S$ or $R \Join S^-$. Otherwise, $a,b,c$ are heavy values so $M_1$ and $M_2$ matrix will contain an entry for $(a,b)$ and $(b,c)$ respectively.}

\introparagraph{Analysis} We now provide a runtime analysis of the above algorithm, and discuss how to optimally choose $\Delta_1, \Delta_2$.

We first bound the running time of the first step. To compute the full join result (before projection), a worst-case optimal algorithm needs time $O(\sNR + \sNS + |\tOUT_{\Join}|)$, where $ |\tOUT_{\Join}|$ is the size of the join. 
The main observation is that the size of the join is bounded by $\sNS \cdot \Delta_1 + |\tOUT| \cdot \Delta_2$. Hence, the running time of the first step is $ O(\sNR + \sNS \cdot \Delta_1 + |\tOUT| \cdot \Delta_2)$.

To bound the running time of the second step, we need to bound appropriately the dimensions of the two rectangular matrices that correspond to the subrelations $R^+,S^+$. Indeed, the heavy $x$-values for $R^+$ are at most ${\sNR}/{\Delta_2}$, while the heavy $y$-values are at most ${\sNS}/{\Delta_1}$. This is because $|\domain(y)| \leq \sNS$. 
Hence, the dimensions of the matrix for $R^+$ are $({\sNR}/{\Delta_2}) \times ({\sNS}/{\Delta_1})$. Similarly, the dimensions of the matrix for $S^+$ are $ ({\sNS}/{\Delta_1}) \times ({\sNS}/{\Delta_2})$. \hlrone{The matrices are represented as two-dimensional arrays and can be constructed in time $C = \max\{{\sNR}/{\Delta_2} \cdot {\sNS}/{\Delta_1},{\sNS}/{\Delta_1}\cdot {\sNS}/{\Delta_2}\}$ by simply iterating over all possible heavy pairs and checking whether they form a tuple in the input relations}. Thus, from~\autoref{lem:matrix:multiplication} the running time of the matrix multiplication step is $M(\frac{\sNR}{\Delta_2}, \frac{\sNS}{\Delta_1},\frac{\sNS}{\Delta_2})$.
Summing up the two steps, the cost of the algorithm is in the order of:
\begin{align} \label{eq:cost}
\sNR + \sNS \ \Delta_1 + |\tOUT| \Delta_2 + M\big(\frac{\sNR}{\Delta_2}, \frac{\sNS}{\Delta_1},\frac{\sNS}{\Delta_2}\big) + \hlrone{ C}
\end{align} 

Using the above formula, one can plug in the formula for the matrix multiplication cost and solve to find the optimal values for $\Delta_1, \Delta_2$. We show how to do this in Section~\ref{sec:optimizer}. 

In the next part, we provide a theoretical analysis for the case where matrix multiplication is achievable with the theoretically optimal $\omega = 2$ for the case where $N_R = N_S = N$. \hlrone{Observe that the matrix construction cost $C$ is of the same order as $M(\frac{\sNR}{\Delta_2}, \frac{\sNS}{\Delta_1},\frac{\sNS}{\Delta_2})$ even when $\omega = 2$, since $\beta$ is the smallest of the three terms ${\sNR}/{\Delta_2}, {\sNS}/{\Delta_1}, {\sNS}/{\Delta_2}$}. Thus, it is sufficient to minimize the  expression
\begin{align*}
f(\Delta_{1}, \Delta_{2}) = N + N \cdot \Delta_1 + |\tOUT| \cdot \Delta_2 + \frac{N^2}{\Delta_2 \min \{ \Delta_1, \Delta_2\}}
\end{align*} 
while ensuring $1 \leq \Delta_{1}, \Delta_{2} \leq N$. \eat{To achieve this, we need to equate the last three terms (since two are increasing in $\Delta_1, \Delta_2$ and the third decreasing). We distinguish two cases. We distinguish two cases.
	
	\smallskip
	\introparagraph{Case 1} $|\tOUT| \leq N$. To minimize the running time we equate $N \cdot \Delta_1 = |\tOUT| \cdot \Delta_2 = N^2/ \Delta_2 \Delta_1$, it holds that  In this case, the minimizer of the expression will have to satisfy $\Delta_2 \geq \Delta_1$, so we have . Solving this system of 
	equations gives us $\Delta_1 = |\tOUT|^{1/3}$, $\Delta_2 = N / |\tOUT|^{2/3}$, and running time
	\[ N + N \cdot |\tOUT|^{1/3} \]
	
	\smallskip
	\introparagraph{Case 2} $|\tOUT| \geq N$. In this case, the minimizer of the expression will have to satisfy $\Delta_2 \leq \Delta_1$, so we have $N \cdot \Delta_1 = |\tOUT| \cdot \Delta_2 = N^2/ \Delta_2^2$. Solving this system of 
	equations gives us $\Delta_1 = |\tOUT|^{2/3}/N^{1/3}$, $\Delta_2 = N^{2/3} / |\tOUT|^{1/3}$, and running time
	\[ N + N^{2/3} \cdot |\tOUT|^{2/3} \]
}

\hlrone{
	The first observation is that for any feasible solution $f(x, y)$ where $x > y$, we can always improve the solution by decreasing the value of $\Delta_{1}$ from $x$ to $y$. Thus, w.l.o.g. we can impose the constraint $1 \leq \Delta_{1} \leq \Delta_{2} \leq N$. 
	
	\smallskip
	\introparagraph{Case 1} $|\tOUT| \leq N$. Since $\Delta_{1} \leq \Delta_{2}$, we have $f(\Delta_{1}, \Delta_{2}) = N \cdot \Delta_1 + |\tOUT| \cdot \Delta_2 + {N^2}/{\Delta_2 \cdot \Delta_1}$. To minimize the running time we equate ${\partial f}/{\partial \Delta_{1}} = N - {N^2}/(\Delta_2 \Delta_{1}^2) = 0$ and ${\partial f}/{\partial \Delta_{2}} = \tOUT - {N^2}/(\Delta_1 \Delta_{2}^2) = 0$. Solving this system of equations gives that the critical point has $\Delta_1 = |\tOUT|^{1/3}$, $\Delta_2 = N / |\tOUT|^{2/3}$. Since $|\tOUT| \leq N$, this solution is feasible, and it can be verified that it is the minimizer of the running time, which becomes 
	\[ N + N \cdot |\tOUT|^{1/3} \]

	\smallskip
	\introparagraph{Case 2} $|\tOUT| > N$. For this case, there is no critical point inside the feasible region, so we will look for a minimizer at the border, where $\Delta_{1} = \Delta_{2} = \Delta$. This condition gives us $f(\Delta) = (N + |\tOUT|) \cdot \Delta + {N^2}/{\Delta^2}$, with minimizer $\Delta = \big( 2N^2/ (N + |\tOUT|) \big)^{1/3}$. The runtime then becomes
	\[ O(N^{2/3} \cdot |\tOUT|^{2/3}) \]
}
We can summarize the two cases with the following result.

\begin{lemma}\label{lem:jp}
	Assuming that the exponent in matrix multiplication is $\omega=2$, the query $\mtwopath$ can be computed in time
	\begin{align*} \label{eq:opt:time}
	O(|\bD| + |\bD|^{2/3} \cdot |\tOUT|^{1/3} \cdot \max \{|\bD|,|\tOUT|  \}^{1/3})
	\end{align*}
\end{lemma}

\autoref{lem:star:basic} implies a running time of $O(|\bD| \cdot |\tOUT|^{1/2})$ for $\mtwopath$, which is strictly worse compared to the running time of the above lemma {\em for every output size} $|\tOUT|$.

\hlrone{
	\smallskip
	\introparagraph{Remark} For the currently best known value of $\omega = 2.37$, the running time is $O(|\bD|^{0.83} \cdot |\tOUT|^{0.589} + |\bD| \cdot |\tOUT|^{0.41})$. 
}

\hlrtwo{
	\smallskip
	\introparagraph{Optimality} The algorithm is worst-case optimal (up to constant factor) for $|\tOUT| = \Theta(N^2)$. The running time becomes $O(N^2)$ which matches the lower bound $|\tOUT|$, since we require at least that much time to enumerate the output.
}

\hlrone{
	\smallskip
	\introparagraph{Comparing with previous results} We now discuss the result in~\cite{amossen2009faster}, which uses matrix multiplication to give a running time of $\tilde{O}(|\bD|^{0.862} \cdot |\tOUT|^{0.408} + |\bD|^{2/3} \cdot |\tOUT|^{2/3})$. We point out an error in their analysis that renders their claim incorrect for the regime where $|\tOUT| < N$.
	
	In order to obtain their result, the authors make a split of tuples into light and heavy, and obtain a formula for running time in the order of $N\Delta_b + |\tOUT|\Delta_{ac} + M\big(\frac{N}{\Delta_{ac}}, \frac{N}{\Delta_b},\frac{N}{\Delta_{ac}}\big)$, where $\Delta_b, \Delta_{ac}$ are suitable degree thresholds. Then, they use the formula from~\cite{huang1998fast} for the cost of matrix multiplication, where $M(x,y,x) = x^{1.84} \cdot y^{0.533} + x^2$. However, this result can be applied only when $x \geq y$, while the authors apply it for regimes where $x < y$. (Indeed, if say $x = N^{0.3}$ and  $y = N^{0.9}$, then we would have $M(x,y,x) = N^{1.03}$, which is smaller than the input size $N^{1.2}$.) Hence, the running time analysis is valid only when $N/\Delta_{ac} \geq N / \Delta_b$, or equivalently $\Delta_b \geq \Delta_c$. Since the thresholds are chosen such that $N\Delta_b = |\tOUT|\Delta_{ac}$, it means that the result is correct {\em only in the regime where $|\tOUT| \geq N$.} In other words, when the output size is smaller than the input size, the running time formula from~\cite{amossen2009faster} is not applicable.
	
	In the case where $\omega = 2$, the cost formula from~\cite{huang1998fast} becomes $M(x,y,x) = x^2$, and \cite{amossen2009faster} gives an improved running time of $\tilde{O}(N^{2/3} \cdot |\tOUT|^{2/3})$. Again, this is applicable only when $|\tOUT| \geq N$, in which case it matches the bound from Lemma~\ref{lem:jp}. Notice that for $|\tOUT| < N^{1/2}$ the formula would imply a deterministic sublinear time algorithm. 

}

\subsection{The Star Query}

We now generalize the result to the star query $Q^\star_k$. As before, we assume that all tuples that do not contribute to the join output have already been removed.

\smallskip

\introparagraph{Algorithm}
The algorithm is parametrized by two integer constants $\Delta_1, \Delta_2 \geq 1$. We partition each relation $R_i$ into three parts, $R^+_i, R^-_i$ and $R^\diamond_i$:
\begin{align*}
R^-_i & = \{R_i(a,b) \mid |\sigma_{x_i=a} R_i(x_i,y)| \leq \Delta_2 \}  \\
R^\diamond_i & = \{R_i(a,b) \mid |\sigma_{y=b} R_j(x_j, y)| \leq \Delta_1, \text{ for each } j \in [k] \setminus i \} \\
R_i^+ & = R_i \setminus (R_i^- \cup R^\diamond_i ) 
\end{align*}
In other words, $R^-_i$ contains all tuples with light $x$, $R^\diamond_i$ contains all tuples with $y$ values that are light in all other relations, and $R_i^+$ the remaining tuples. The algorithm now proceeds by computing the following result:

\begin{enumerate}
	\item \label{eq1} Compute $R_1 \Join \dots R^-_j \Join \dots R_k$ using any worst-case optimal join algorithm, then project for each $j \in [k]$. 
	\item \label{eq2} Compute $R_1 \Join \dots R^\diamond_j \Join \dots R_k$ using any worst-case optimal join algorithm, then project for each $j \in [k]$. 
	\item Materialize $R^+_1, \dots, R^+_k$ as rectangular matrices and use matrix multiplication to compute their product. 
\end{enumerate}

\introparagraph{Analysis}
We assume that all relation sizes are bounded by $N$.
The running time of the first step is $O(|\tOUT| \cdot \Delta_{2})$ since each light value of variable $x_i$ in relation $R_i$ contributes to at least one output result.

For the second step, the key observation is that since $y$ is light in all relations (except possibly $R_i$), the worst-case join size before projection is bounded by $O(\sN \cdot \Delta_{1}^{k-1})$, and hence the running time is also bounded by the same quantity. 

The last step is more involved than simply running matrix multiplication. This is because for each output result formed by heavy $x_i$ values in $R^+_i$ (say $t = (a_1, a_2, \dots a_k)$), we need to count the number of $y$ values that connect with each $a_i$ in $t$. However, running matrix multiplication one at a time between two matrices only tells about the number of connection $y$ values for any two pair of $a_i$ and not all of $t$. In order to count the $y$ values for all of $t$ together, we divide  variables $x_1, \dots x_k$ into two groups of size $\lceil k/2 \rceil$ and $\lfloor k/2 \rfloor$ followed by creating two adjacency matrices. Matrix $V$ is of size $\big(\frac{\sN}{\Delta_2}\big)^{\lceil k/2 \rceil} \times \frac{\sN}{\Delta_1}$ such that 
\begin{align*}
V_{(a_1, a_2, \dots a_{\lceil k/2 \rceil}),b} = \begin{cases}
1, & (a_1, b) \in {R}_1, \dots, (a_{\lceil k/2 \rceil}, b) \in {R}_{{\lceil k/2 \rceil}}  \\
0, & \text{otherwise}
\end{cases}
\end{align*}
Similarly, matrix $W$ is of size $ \big(\frac{\sN}{\Delta_2}\big)^{\lfloor k/2 \rfloor} \times \frac{\sN}{\Delta_1}$ such that 
\begin{align*}
W_{(a_{\lceil k/2 \rceil + 1} \dots a_{k}),b} = \begin{cases}
1, & (a_{\lceil k/2 \rceil + 1}, b) \in {R}_{\lceil k/2 \rceil + 1}, \dots, (a_{k}, b) \in {R}_{k}  \\
0, & \text{otherwise}
\end{cases}
\end{align*}

\begin{example}
	\hlrtwo{Consider the relations $R(x,y)$ and $S(z,y)$ from previous example and consider relation $T(p,y)$ and $U(q,y)$ as shown below}. 
	
	\hspace{3em}
	\begin{minipage}{\linewidth}
		\scalebox{.85}{\begin{tikzpicture}
			
			\begin{scope}[fill opacity=1]
			
			\node at (-1,1) {Relation $T$};							
			\node at (-2,0.5) {\color{gray}$\mathbf{p}$};		
			\node at (0,0.5) {\color{orange}$\mathbf{y}$};		
			\node (x1) at (-2,0) {$1$};
			\node (x2) at (-2,-0.5) {$2$};
			\node (x3) at (-2,-1) {$3$};
			\node (x4) at (-2,-1.5) {$4$};
			\node (x5) at (-2,-2) {$5$};
			\node (x6) at (-2,-2.5) {$6$};

			\node (y1) at (0,0) {$1$};
			\node (y2) at (0,-0.5) {$2$};
			\node (y3) at (0,-1) {$3$};
			\node (y4) at (0,-1.5) {$4$};
			\node (y5) at (0,-2) {$5$};
			\node (y6) at (0,-2.5) {$6$};			
			
			\draw[-, color=red] (x1) to (y1);
			\draw[-, color=red] (x1) to (y3);
			\draw[-, color=red] (x2) to (y2);
			\draw[-, color=red] (x6) to (y1);
			\draw[-, color=red] (x3) to (y3);
			\draw[-, color=red] (x3) to (y4);
			\draw[-] (x4) to (y4);
			\draw[-] (x4) to (y5);
			\draw[-] (x4) to (y6);
			\draw[-] (x5) to (y4);
			\draw[-] (x5) to (y5);
			\draw[-] (x5) to (y6);			
			\draw[-, color=red] (x6) to (y2);
			\draw[-] (x6) to (y5);
			\draw[-] (x6) to (y6);						
			
			\node at (3,1) {Relation $U$};										
			\node at (2,0.5) {\color{violet} $\mathbf{q}$};		
			\node at (4,0.5) {\color{orange}$\mathbf{y}$};		
			\node (z1) at (2,0) {$1$};
			\node (z2) at (2,-0.5) {$2$};
			\node (z3) at (2,-1) {$3$};
			\node (z4) at (2,-1.5) {$4$};
			\node (z5) at (2,-2) {$5$};
			\node (z6) at (2,-2.5) {$6$};

			\node (yy1) at (4,0) {$1$};
			\node (yy2) at (4,-0.5) {$2$};
			\node (yy3) at (4,-1) {$3$};
			\node (yy4) at (4,-1.5) {$4$};
			\node (yy5) at (4,-2) {$5$};
			\node (yy6) at (4,-2.5) {$6$};			
			
			\draw[-, color=green] (z1) to (yy1);
			\draw[-, color=green] (z2) to (yy2);
			\draw[-, color=green] (z2) to (yy5);			
			\draw[-, color=green] (z3) to (yy3);
			\draw[-] (z4) to (yy4);
			\draw[-] (z4) to (yy5);
			\draw[-] (z4) to (yy6);
			\draw[-] (z5) to (yy4);
			\draw[-] (z5) to (yy5);
			\draw[-] (z5) to (yy6);			
			\draw[-] (z6) to (yy4);
			\draw[-] (z6) to (yy5);
			\draw[-] (z6) to (yy6);

			\end{scope}	
			\end{tikzpicture}
		}
	\end{minipage}
	
	\hlrtwo{Suppose that we wish to compute the result of star query $Q^\star_4 = R(x,y), S(z, y), T(p, y), U(q, y)$. Similar to the previous example, we fix $\Delta_{1} = \Delta_{2} = 2$. It is easy to verify for this example $R^- = R^\diamond$ and similarly for all other relations. We can now evaluate the join as mentioned in step~(\ref{eq1}) and~(\ref{eq2}). Next, we construct the matrices $V$ and $W$. We divide the variables $x,z,p,q$ into groups, $x,z$ and $p,q$. $V$ will consist of all heavy $(x,z)$ pairs ($9$ in total) as one dimension and $y$ as the second dimension. Similarly, $W$ consists of all heavy $(p,q)$ pairs ($9$ in total) as one dimension and $y$ as the other.}
	
 \begin{tikzpicture}[mine/.style={anchor=base,text depth=0ex,text height=1.5ex,text width=1.5em}, scale=0.7]
	\matrix (K) [matrix of nodes, nodes={mine}, column sep=0mm, row sep=0mm, nodes={draw, fill=blue!10},ampersand replacement=\&]
	{
		\& |[fill=orange!40]|4 \& |[fill=orange!40]|5 \& |[fill=orange!40]|6 \\
		\cab{4}{brown}{4}{cyan} \& 1 \& 1 \& 1 \\
		\cab{4}{brown}{5}{cyan} \& 1 \& 0 \& 1 \\
		$\dots$ \& $\dots$ \& $\dots$ \& $\dots$ \\
		\cab{6}{brown}{6}{cyan} \& 0 \& 1 \& 0 \\
	};
	
	\node[below= 1.4em of K-4-2.south] (l1) {\qquad matrix $M_1$};
	
	\node[right= 0.5em of K-4-3.east] (l2) {\qquad $\times$};

	\matrix (L) [right=1.5em of K,matrix of nodes, nodes = {mine}, column sep=0mm, row sep=0mm, nodes={draw, fill=blue!10},ampersand replacement=\&]
	{
		\& \cab{4}{gray}{4}{violet} \& \cab{4}{gray}{5}{violet} \& $\dots$ \& \cab{6}{gray}{6}{violet} \\
		|[fill=orange!40]|4 \& 1 \& 1 \& $\dots$   \& 0 \\
		|[fill=orange!40]|5 \& 1 \& 1 \& $\dots$   \& 1 \\
		|[fill=orange!40]|6 \& 1 \& 1 \& $\dots$   \& 1 \\
	};
	
	\node[below= 1.5em of L-3-3.south] (l3) {\qquad matrix $M_2$};

	\end{tikzpicture} 
	
\end{example}

Matrix construction takes time $({\sN}{/\Delta_2})^{\lceil k/2 \rceil} \cdot {\sN}/{\Delta_1}$ time in total. We have now reduced step three in computing the matrix product $V \times W^T$.  Summing up the cost of all three steps, the total cost is in the order of
\begin{align*} 
\sN\cdot \Delta_1^{k-1} + |\tOUT| \cdot \Delta_2 + M\big(\big(\frac{\sN}{\Delta_2}\big)^{\lceil k/2 \rceil}, \frac{\sN}{\Delta_1}, \big(\frac{\sN}{\Delta_2}\big)^{\lfloor k/2 \rfloor} \big)
\end{align*}
Similar to the two-path query, we can find the exact value of $\Delta_1$ and $\Delta_2$ that minimizes the total running cost given a cost formula for matrix multiplication. 
We conclude this subsection with an illustrative example to show the benefit of matrix multiplication over the combinatorial algorithm.

\begin{example}
	Let $k=3$ and $|\tOUT| = N^{3/2}$. The running time is minimized when
	$$ N \cdot\Delta_1^{2} = |\tOUT| \cdot \Delta_2 =  M\big( \big(\frac{N}{\Delta_2}\big)^{2}, \frac{N}{\Delta_1}, \frac{N}{\Delta_2}\big) $$
	
	The first equality gives us $\Delta_{1}^2 = \sqrt{N} \cdot \Delta_{2}$. We will choose the thresholds such that, $\Delta_{2} < \Delta_{1}$. This means $\beta = N / \Delta_{1}$. From the second equality, we get $ |\tOUT| \cdot \Delta_2 = \big(\frac{N}{\Delta_2}\big)^ {3}$. The running time is minimized when $\Delta_{2} = N^\frac{6}{16}, \Delta_{1} = \sN^\frac{7}{16}$, in which case it is  $ O(N^\frac{15}{8})$ which is sub-quadratic (assuming $\omega=2$). In contrast,~\autoref{lem:star:basic} has a worse running time $O(N^{\text{ }2})$.
\end{example}

\subsection{Boolean Set Intersection} \label{sec:bsi}

In this setting, we are presented with a workload $W$ containing boolean set intersection (BSI) queries of the form $Q_{ab}() = R(a,y),S(b,y)$ parametrized by the constants $a,b$. The queries come at a rate of $B$ queries/time unit. In order to service these requests, we can use multiple machines. Our goal is twofold: minimize the number of machines we use, while at the same time minimizing the {\em average latency}, defined as the average time to answer each query.

\begin{example} \label{ex:bsi}
	The simplest strategy is to answer each request using a separate machine. Computing a single BSI query takes worst-case time $O(N)$, where $N$ is the input size. Hence, the average latency is $O(N)$. At the same time, since queries come at a rate of $B$ queries per time unit, we need $\rho = B \cdot N$ machines to keep up with the workload.
\end{example}

Our key observation is that, instead of servicing each request separately, we can {\em batch} requests and compute them all at once. To see why this can be beneficial, suppose that we batch together $\cC$ queries. Then, we can group all pairs of constants $(a,b)$ to a single binary relation $T(x,z)$ of size $\cC$, and compute the following conjunctive query:
\[ Q_{batch}(x,z) = R(x,y),S(z,y),T(x,z). \]
Here, $R,S$ have size $N$, and $T$ has size $\cC$. The resulting output will give the subset of the pairs of sets that indeed intersect. The above query can be computed by applying a worst-case optimal algorithm and then performing the projection: this will take $O(N \cdot \cC^{1/2})$ time. Hence, the average latency for a request will be $\frac{\cC}{B} + \frac{N}{\cC^{1/2}}$. 

To get even lower latency, we can apply a variant of the AYZ algorithm~\cite{alon1997finding} that uses fast matrix multiplication. The algorithm works as follows. For $x,y,z$ values with degrees less than some threshold $\Delta \geq 1$, we perform the standard join with running time $O(N \cdot \Delta)$. For the remaining values, we express relations $R,S$ as rectangular matrices of dimensions $\frac{\cC}{\Delta} \times \frac{N}{\Delta}$ and $\frac{N}{\Delta} \times \frac{\cC}{\Delta}$ respectively. We compute the matrix product, and then intersect the result with $T$ to obtain the final output. The running time for this step is $M (\frac{\cC}{\Delta}, \frac{N}{\Delta}, \frac{\cC}{\Delta})$, which for $\omega=2$ becomes $O(S \cdot N/ \Delta^2)$. To minimize the running time for both steps, we choose $\Delta = \cC^{1/3}$, and thus the running time becomes $O(N \cdot \cC^{1/3})$.

Using the above algorithm, we can show the following.

\begin{proposition} \label{lem:wcoj}
	Let $W$ be a workload of queries of the form $Q_{ab}() = R(a,y),S(b,y)$ at the rate of $B$ access requests per second. Then, assuming the exponent of matrix multiplication is $\omega=2$, we can achieve an average latency of $O(N^{3/5} / B^{2/5})$ using $\rho =  (B \cdot N)^{3/5}$ machines.
\end{proposition}

\begin{proof}
	Using batch size $\cC$, we can process $\cC$ queries in time $O(N \cdot \cC^{1/3})$. Hence, the average latency in this case is
	$O \left( \frac{N}{\cC^{2/3}} + \frac{{\cC}}{\textsf{B}} \right)$.
	To minimize the latency, we choose $\cC = (B \cdot N)^{3/5}$, in which case we obtain an average latency of $O(N^{3/5} / B^{2/5})$.
	Then, the number of machines needed to service the workload is $(B \cdot N) / \cC^{2/3} = (B \cdot N)^{3/5}$.
\end{proof}

Observe that the above proposition strictly improves the number of machines compared to the baseline approach of~\autoref{ex:bsi}. However, the average latency is smaller only for $B \leq N^{3/2}$, otherwise it is larger.

In our experiments, we use the requests in the batch to filter the relations $R$ and $S$, and then compute the 2-path query using~\autoref{algo:two:path}.

\section{Speeding Up \textsf{SSJ} and \textsf{SCJ}} \label{sec:application}

In this section, we will see how to use~\autoref{algo:two:path} to speed up \textsf{SSJ} and \textsf{SCJ}.

\begin{algorithm} [!t]
	\SetCommentSty{textsf}
	\DontPrintSemicolon 
	\SetKwFunction{fullreducer}{\textsc{Materialize}}
	\SetKwFunction{initializepq}{\textsc{InitializePQ}}
	\SetKwFunction{recurse}{\textsc{Recurse}}
	\SetKwFunction{proc}{\textsf{eval}}
	\SetKwFunction{ins}{\textsc{insert}}
	\SetKwFunction{app}{\textsc{insert}}
	\SetKwFunction{top}{\textsc{top()}}
	\SetKwFunction{GetSizeBoundary}{\textsc{GetSizeBoundary}}
	\SetKwData{true}{\textsf{true}}
	\SetKwData{light}{$\mathsf{t_{light}}$}
	\SetKwData{heavy}{$\mathsf{t_{heavy}}$}
	\SetKwData{prev}{$\mathsf{prev}$}
	\SetKwData{clight}{$\mathsf{cost_{light}}$}
	\SetKwData{cheavy}{$\mathsf{cost_{heavy}}$}
	\SetKwData{cdf}{$\mathsf{cdf}$}
	\SetKwFunction{fillout}{\textsc{FillOUT}}
	\SetKwInOut{return}{\textsc{return} \noindent}
	\SetKwInOut{Output}{\textsc{output} \noindent}
	
	\KwIn{ Indexed sets $R = \{R_1, \dots, R_m\}$ and $c$}
	\KwOut{Unordered \textsf{SSJ} result}
	\BlankLine
	\textit{degree threshold }$x = \GetSizeBoundary(R, c), \mL \leftarrow \emptyset$ \;
	$R = R_l \cup R_h$  \tcc*{partition sets into light and heavy}
	\textit{Evaluate $R \Join R_h$ and enumerate result} \label{line:heavy} \;
	\ForEach{$r \in R_l$}{
		\ForEach{$c$\textit{-subset }$r_c$ of $r$}{
			$\mL[r_c] = \mL[r_c] \cup r$ \;	
		}
	}
	\ForEach{$l \in \mL[r_c]$} {
		\textit{enumerate every set pair in $l$ if not output already} \label{line:light} \;
	}
	\caption{\texttt{SizeAware}~\cite{deng2018overlap}}
	\label{algo:ssj}
\end{algorithm}

\smallskip
\noindent \introparagraph{Unordered \textsf{SSJ}} We will first briefly review the state-of-the-art algorithm from~\cite{deng2018overlap} called \texttt{SizeAware}. Algorithm~\ref{algo:ssj} describes the size-aware set similarity join algorithm. The key insight is to identify degree threshold $x$ and partition the input sets into light and heavy. All heavy sets that form an output pair are enumerated by a sort merge join. All light sets are processed by generating all possible $c$-sized subsets and then building an inverted index over it that allows for enumerating all light output pairs. $x$ is chosen such that the cost of processing heavy and light sets is equal to each other. 

We propose three key modifications that give us the new algorithm called \texttt{SizeAware++}. First, observe that $\mathcal{J}_H = R \Join R_h$ (\autoref{line:heavy}) is a natural join and requires $\sN \cdot \sN / x$ operations (recall that $|R_h| = \sN/x$ in the worst-case) even if the join output is smaller. Thus,~\autoref{algo:two:path} is applicable here directly. This strictly improves the theoretical worst-case complexity of~\autoref{algo:ssj} whenever $|\mathcal{J}_H| < \sN^2/x$ for all $c$. 

The second modification is to deal with high duplication when enumerating all light pairs using the inverted index $\mL[r_c]$. The key observation is that~\autoref{line:light} is also performing a brute-force operation by going over all possible pairs and generating the full join result. This step takes $|\mathcal{J}_L| = \sum_{r_c} |\mL[r_c]|^2$ time. If the final output is smaller than $|\mathcal{J}_L|$, then we can do better by using matrix multiplication based algorithm.

\def\lshift{-3}
\def\rshift{1.5}
\begin{figure}
	\scalebox{0.8}{
		\begin{tikzpicture}
		\node at (-2.75+\lshift,0) {$A_1$};
		\node[rectangle, draw, minimum width=7mm] at (-1+\lshift,0) {\hspace{0.1em} $b_1$ \hspace{0.9em} $b_2$ \hspace{0.9em} $b_3$};
		\node[rectangle, draw, minimum width=4mm] at (-1.8+\lshift,-1.6) {\shortstack{$C_1$ \vspace*{0.7em} \\  $C_2$\vspace*{0.7em} \\  $C_3$\vspace*{0.7em} \\  $C_4$}};
		\node[rectangle, draw, minimum width=4mm] at (-0.95+\lshift,-1.3) {\shortstack{$C_1$ \vspace*{0.7em} \\  $C_2$\vspace*{0.7em} \\  $C_3$}};
		\node[rectangle, draw, minimum width=4mm] at (-0.15+\lshift,-0.98) {\shortstack{  $C_3$\vspace*{0.7em} \\  $C_5$}};

		\node at (-2.75+\rshift,0) {$A_2$};
		\node[rectangle, draw, minimum width=7mm] at (-1+\rshift,0) {\hspace{0.1em} $b_1$ \hspace{0.9em} $b_2$ \hspace{0.9em} $b_4$};
		\node[rectangle, draw, minimum width=4mm] at (-1.8+\rshift,-1.6) {\shortstack{$C_1$ \vspace*{0.7em} \\  $C_2$\vspace*{0.7em} \\  $C_3$\vspace*{0.7em} \\  $C_4$}};
		\node[rectangle, draw, minimum width=4mm] at (-0.95+\rshift,-1.3) {\shortstack{$C_1$ \vspace*{0.7em} \\  $C_2$\vspace*{0.7em} \\  $C_3$}};
		\node[rectangle, draw, minimum width=4mm] at (-0.15+\rshift,-0.98) {\shortstack{  $C_4$\vspace*{0.7em} \\  $C_6$}};
		
		\end{tikzpicture}}
	\caption{Example instance showing inverted lists} \label{fig:inverted}
\end{figure}

\begin{figure} 
	\captionsetup[subfigure]{font=footnotesize}
	\centering
	\subcaptionbox{Before materialization}[0.42\linewidth]{
		\scalebox{0.8}{\begin{tikzpicture}
			\tikzset{
				treenode/.style = {align=center, inner sep=0pt, text centered},
				arn/.style = {treenode, circle, black, draw=black,
					fill=white, text width=3.1ex},
				arnrec/.style = {treenode, rectangle, black, draw=black,
					fill=white, text width=4.5ex,minimum width=4.0ex, minimum height=4.0ex},
				arnsmall/.style = {treenode, circle, black, draw=black,
					fill=white, text width=4.5ex},
			}
			\node [sibling distance=10cm,level distance = 4cm]  [arn] {$b_1$}
			child{ [sibling distance=0.8cm, black] node (A) [arn] {$b_2$}
				child{[sibling distance=0.6cm,level distance = 1cm, black]node [arn] {$b_3$}
					child{[black] node[arnrec] {$A_1$}
					}
				}
				child{[sibling distance=0.6cm,level distance = 1cm, black] node [arn] {$b_4$}
					child{[black] node (C) [arnrec] {$A_2$}
					}
				}
			}
			child{ [sibling distance=0.8cm, level distance=1.5cm, black] node (B) [arn] {$b_5$}
				child{[sibling distance=0.6cm,level distance = 1cm, black] node [arn] {$b_7$}
					child{[black] node[arnrec] {$A_3$}
					}
				}
				child{[black] node[arnrec] {$A_4$}
				}
			};
			\node [left=of A, xshift = 3em, red] { \scriptsize $U_1$};
			\node [right=of B, xshift = -3em, blue] { \scriptsize $U_2$};
			\node [below=of C, yshift = 3em, red] (D) { \scriptsize $U_1 = \emptyset$};
			\node [below=of D, yshift = 3em, blue] { \scriptsize $U_2 = \emptyset$};
			\end{tikzpicture}}}%
	\subcaptionbox{After materialization}[0.42\linewidth]{
		\scalebox{0.8}{\begin{tikzpicture}
			\tikzset{
				treenode/.style = {align=center, inner sep=0pt, text centered},
				arn/.style = {treenode, circle, black, draw=black,
					fill=white, text width=3.1ex},
				arnrec/.style = {treenode, rectangle, black, draw=black,
					fill=white, text width=4.5ex,minimum width=4.0ex, minimum height=4.0ex},
				arnsmall/.style = {treenode, circle, black, draw=black,
					fill=white, text width=4.5ex},
			}
			\node [sibling distance=10cm,level distance = 4cm]  [arn] {$b_1$}
			child{ [sibling distance=0.8cm, black] node (A) [arn] {$b_2$} 
				child{[sibling distance=0.6cm,level distance = 1cm, black]node [arn] {$b_3$}
					child{[black] node[arnrec] {$A_1$}
					}
				}
				child{[sibling distance=0.6cm,level distance = 1cm, black] node [arn] {$b_4$}
					child{[black] node (C) [arnrec]  {$A_2$}
					}
				}
			}
			child{ [sibling distance=0.8cm, level distance=1.5cm, black] node (B) [arn] {$b_5$}
				child{[sibling distance=0.6cm,level distance = 1cm, black] node [arn] {$b_7$}
					child{[black] node[arnrec] {$A_3$}
					}
				}
				child{[black] node[arnrec] {$A_4$}
				}
			};
			\node [left=of A, xshift = 3em, red] { \scriptsize $U_1$};
			\node [right=of B, xshift = -3em, blue] { \scriptsize $U_2$};
			\node [below=of C, yshift = 3em, red] (D) { \scriptsize $U_1 = \{C_4\}$};
			\node [below=of D, yshift = 3em, blue] { \scriptsize $U_2 = \{C_1, C_2\}$};
			\end{tikzpicture}}}
	\label{fig:runexample}	
	\caption{Materialization in prefix tree}
\end{figure}

The final observation relates to optimizing the expansion of light nodes (\autoref{line:first} in~\autoref{algo:two:path}). Recall that the algorithm expands all light $y$ values. Suppose we have ${R(x,y)}$ and ${S(z,y)}$ relations indexed and sorted according to variable order in the schema. Let $\mL[b] = \setof{c}{ (c,b) \in S(z,y)}$ denote the inverted index for relation $S$. The time required to perform the deduplication for a fixed value for $x$ (say $a$) is $T = \sum_{b : (a,b) \in R} |\mL[b]|$. This is unavoidable for overlap $c=1$ in the worst case. However, it is possible that for $c>1$, the output size is much smaller than $T$. In other words, deduplication step is expensive when the overlap between $\mL[b]$ for different $b$ is high. The key idea is to reuse computation across multiple $a$ if there is a shared prefix and high overlap. We illustrate the key idea with the following example.

\begin{example} \label{ex:ssj}
	Consider the instance shown in~\autoref{fig:inverted} with $\{b_1, \dots, b_7\}$ as the possible keys for inverted index $\mL[b]$ and sets $A_1,\dots A_4$ as shown in the prefix tree. We use the length of inverted list $\mL[b]$ as the key for sorting the input sets descending order. Suppose overlap $c=2$. After merging inverted list for $b_1, b_2$, we know that $C_1, C_2, C_3$ are present at least two times across all of the lists. At this point, we materialize two things: $\textsf{(i)}$ $O_1 = \{C_1, C_2, C_3\}$ as output and, $\textsf{(ii)}$ $U_1 = \mL[b_1] \cup \mL[b_2] \setminus O_1$ at the node $b_2$ in the tree. We then continue with merging the other lists. When we start the enumeration for $A_2$, we know that the lists $b_1, b_2$ have already been processed and we can simply use the stored output. For $b_4$, we can go over its content and check whether the value is present in $U_1$. Similarly, for sets $A_3$ and $A_4$ that share $b_1, b_5$ as a prefix, $U_2$ stores the union of $\mL[b_1]$ and $\mL[b_5]$ after removing the sets that have appeared at least two times. For $A_1, A_2$, simply performing a merge step requires $9 + 9=18$ operations in total. However, if reusing the computation, we require only $9$ operations: $7$ for $A_1$ (merging $b_1$ and $b_2$) and $2$ for merging inverted list of $b_4$ with stored $U_1$. 
\end{example}

The global sort order for all $b$ is the length of inverted list $\mL[b]$. This encourages more computation reuse since bigger lists will give larger output and merging those repeatedly is expensive. Since a global sort order has been defined, we can construct a prefix tree and store the output and list union  in the prefix tree. This technique will provide the largest gain when input sets $A_i$ have a significant overlap. There also exists a tradeoff between space requirement and computation reuse. Storing the output and list union at every node in the prefix tree increases the materialization requirement. The space usage can be controlled by limiting the depth at which the output and list union is stored. This can avoid excessive materialization when space is limited. The three optimizations in \texttt{SizeAware++} together can deliver speedups up to an order of magnitude over \texttt{SizeAware} for single threaded implementation. Next, we highlight the important aspects regarding parallelization of \textsf{SSJ} and why \texttt{SizeAware} is not as amenable to parallelization as we would it to be. Partitioning join $\mathcal{J}_H$ is straightforward since all parallel tasks require no synchronization and access the input data in a read-only manner. Parallelizing $\mathcal{J}_L$ in \textsf{SizeAware} is harder because of two reasons: $\textsf{(i)}$ generating the $\textsf{c}$-sized subsets requires coordination since a given subset can be generated by multiple small sets; $\textsf{(ii)}$ once the subsets have been generated, a given output pair $(\textsf{r}, \textsf{s})$ can be connected to multiple $\textsf{c}$-subsets. This means that each parallel task needs to coordinate in order to deduplicate multiple results across different $\textsf{c}$-subsets. On the other hand, using matrix multiplication allows for coordination-free parallelism as the matrix can be partitioned easily and each parallel task requires no interaction with each other. We show how this is achieved in~\autoref{sec:implementation}.  Note that \texttt{SizeAware++} also suffers from the same drawback as it is also generating the $c$-sized subsets which can be expensive. For dense datasets, using matrix multiplication and filtering the join result to find the similar set pairs is the fastest technique and also benefits the most from parallelization.

\smallskip
\noindent \introparagraph{Ordered \textsf{SSJ}} In this part, we look at the problem of enumerating SSJ in decreasing order of set similarity. Ordered enumeration of output pairs can be done by first generating the output and then sorting it. Note that the processing of light sets in~\autoref{algo:ssj} (and consequently \texttt{SizeAware++}) is not amenable to finding the set pair with the largest intersection. Once an output pair has been identified, we still need to enumerate over elements in the sets to identify the exact intersection size. On the other hand, our matrix multiplication based join provides with a count that can be used for sorting.

\smallskip
\noindent \introparagraph{\textsf{SCJ}} \textsf{SCJ} algorithms~\cite{bouros2016set} typically prune away most of the set pairs that are surely not contained within each other. This acts as a blocking filter. For the remaining set pairs, the verification step performs a sort-merge join to verify if containment holds for either of the sets i.e we perform a merge join for all set pairs that need to be verified. However, the verification step can be slow if the overlap between sets is high (because of multiple replicas) or the average inverted index size is large. For these cases, we can get a significant speedup by simply evaluating the join-project result. This approach is most beneficial when the set containment join result is close to the join-project result. Further, majority of \textsf{SCJ} algorithms do not use the power of parallel computation. \textsf{PIEJoin}~\cite{kunkel2016piejoin} is the first and the only algorithm that addresses parallel \textsf{SCJ}. Since join processing is highly parallelizable, computing \textsf{SCJ} via join-project output benefits from parallel computation as well.

\section{Cost-Based Optimization} \label{sec:optimizer}

In this section, we deep dive into the challenges of making our framework practical and how we can fine tune the knobs to minimize the running time.

\smallskip
\introparagraph{Estimating output size} So far, we have not discussed how to estimate for $|\tOUT|$. We derive an estimate in the following manner. First, it is simple to show that the following holds for $\mtwopath$: $ |\domain(x)| \leq |\tOUT| \leq \min\{|\domain(x)|^2|,  |\tOUT_{\Join}|\}|$ and $|\tOUT_{\Join}| \leq \mathsf{N} \cdot \sqrt{|\tOUT|}$. Thus, a reasonable  estimate for $|\tOUT|$ is the geometric mean of $\max\{|\domain(x)|, (|\tOUT_{\Join}| / \mathsf{N})^2\}$ and $\min\{|\domain(x)|^2|,  |\tOUT_{\Join}|\}|$. If $|\tOUT_{\Join}|$ is not \emph{much larger} than $\mathsf{N}$, then the full join size is also reasonable estimate. We return to this point later in the discussion of the optimizer.

\smallskip
\introparagraph{Indexing relations}  Join processing by applying worst-case optimal join algorithms is possible only if all relations are indexed over the variables. This means each relation will be stored once for every index order that is required. For a binary relation $R(x,y)$, this would mean storing the relation indexed by all values of $x$ as key and a sorted list of values for $y$ and vice-versa. This can be accomplished in $O(|\bD| \log|\bD|)$ time after removing all tuples that do not join. During this pass, it is also straightforward to compute the size of full join result (i.e., before the projection). Additionally, we create the following indexes:

\begin{enumerate}
	\item For variable $x$ and degree threshold $\delta$, an index that tells us the deduplication effort when performing set union (i.e $\sum_{\text{light } a}\sum_{b : (a,b) \in R} |\mL[b]|$) for all values of $x$ with degree $\leq \delta$. We call this index \textsf{sum}$(x_\delta)$. Similarly for all values of $y$ with degree $\leq \delta$, \textsf{sum}$(y_\delta) = \sum_{\text{light }b} |\mL[b]|^2$.
	\item For the projected out variable $y$ and degree threshold $\delta$, an index that counts the number of $x$ connected to all $y$ values with degree $\leq \delta$. We call this index \textsf{cdfx}$(y_\delta)$.
	\item For each variable (say $w$), an index that tells us the number of values for $w$ with degree $\leq \delta$. We call this index \textsf{count}$(w_\delta)$.
\end{enumerate}

All indexes can be built in linear time by storing the sorted vector containing the true distribution of values present in the relation. Then, given a $\delta$, we can binary search over the vector to find the exact count (sum).

\begin{figure}
	\centering
	\captionsetup[subfigure]{oneside}
	\subcaptionbox{Single core scalability \label{fig:mmul:scale}}%
	[10em]{\hspace{-3.6em}\includegraphics[scale=0.65]{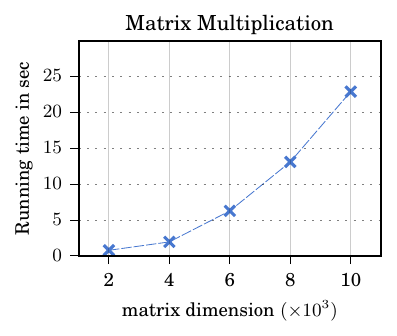}}
	\hspace{0.5em}
	\subcaptionbox{Multi core scalability \label{fig:mmul:parallel}}%
	[10em]{\hspace{-2.5em}\includegraphics[scale=0.65]{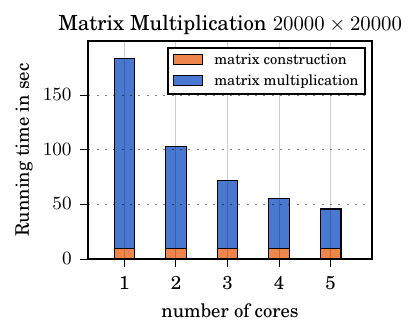}}
	\caption{Matrix Multiplication Running Time}	
\end{figure} 

\smallskip
\introparagraph{Matrix multiplication cost} A key component of all our techniques is matrix multiplication.~\autoref{lem:matrix:multiplication} states the complexity of performing multiplication and also includes the cost of creating the matrices. However, in practice, this could be a significant overhead both in terms of memory consumption and time required. Further, the scalability of matrix multiplication implementation itself is subject to matrix size, the underlying linear algebra framework and hardware support (vectorization, SIMD instructions, multithreading support etc.)
In order to minimize the running time, we need to take into consideration all system parameters in order to estimate the optimal threshold $\Delta$ values.

\begin{table}[!htp]
	\small
	\centering
	\scalebox{1}{
		\begin{tabular}{p{1.5cm}lp{0.4cm}}
			\toprule
			\textbf{Symbol}                    &{\textbf{Description}}  \\ 
			\midrule 
			$T_s$ & avg time for sequential access in $\oldtt{std::vector}$ \\
			$T_m$ & avg time for allocating $32$ bytes of memory  \\
			$co$ & number of cores available \\
			$\hat{M}(u,v,w,co)$ &\begin{minipage}[t]{0.73\columnwidth} estimate of time required to multiply matrices of dimension $u \times v$ and $v \times w$ using $co$ cores\end{minipage} \vspace{0.1em}\\
			$T_I$ & \begin{minipage}[t]{0.73\columnwidth} avg time for random access and insert in $\oldtt{std::vector}$ \end{minipage} \\
			\bottomrule
		\end{tabular}
	}
	\caption{Symbol definitions.}
	\label{tab:symbols}
\end{table}
\vspace{-1em}

\autoref{algo:optimizer} describes the cost based optimizer used to find the best degree thresholds in order to minimize the running time. To simplify the description, we describe the details for the case of $\mtwopath$ where $R = S$ (i.e., a self join). If the full join result is not much larger than the size of input relation, then we can simply use any worst case optimal join algorithm. For our experiments, we set the upper bound for $|\tOUT_{\Join}|$ to be at most $20 \cdot N$. Beyond this point, we begin to see the benefit of using matrix multiplication for join-project computation.

\begin{algorithm} [!t]
	\SetCommentSty{textsf}
	\DontPrintSemicolon 
	\SetKwFunction{fullreducer}{\textsc{Materialize}}
	\SetKwFunction{initializepq}{\textsc{InitializePQ}}
	\SetKwFunction{recurse}{\textsc{Recurse}}
	\SetKwFunction{proc}{\textsf{eval}}
	\SetKwFunction{ins}{\textsc{insert}}
	\SetKwFunction{app}{\textsc{insert}}
	\SetKwFunction{top}{\textsc{top()}}
	\SetKwFunction{pop}{\textsc{pop()}}
	\SetKwData{true}{\textsf{true}}
	\SetKwData{light}{$\mathsf{t_{light}}$}
	\SetKwData{heavy}{$\mathsf{t_{heavy}}$}
	\SetKwData{prevl}{$\mathsf{prev_{light}}$}
	\SetKwData{prevh}{$\mathsf{prev_{heavy}}$}
	\SetKwData{clight}{$\mathsf{cost_{light}}$}
	\SetKwData{cheavy}{$\mathsf{cost_{heavy}}$}
		\SetKwData{done}{$\mathsf{prev_{\Delta_{1}}}$}
	\SetKwData{dtwo}{$\mathsf{prev_{\Delta_{2}}}$}
	\SetKwData{sum}{$\mathsf{sum}$}
	\SetKwData{count}{$\mathsf{count}$}
	\SetKwData{countx}{$\mathsf{cdfx}$}	
	\SetKwFunction{fillout}{\textsc{FillOUT}}
	\SetKwInOut{return}{\textsc{return} \noindent}
	\SetKwInOut{Output}{\textsc{output} \noindent}
	\KwOut{ degree threshold $\Delta_1, \Delta_2$}
	\BlankLine
	\textit{Estimate full join result $|\tOUT_{\Join}|$ and $|\tOUT|$}\;
	\If{$|\tOUT_{\Join}| \leq 20 \cdot N$ }{\textit{use worst-case optimal join algorithm}}
	\BlankLine
	$\light \leftarrow |\tOUT_{\Join}|, \heavy \leftarrow 0, \prevl \leftarrow \infty, \prevh \leftarrow 0, \Delta_1 = \sN$ \;
	
	\While{\true}{
		$\prevl \leftarrow \light, \prevh \leftarrow \heavy$ \;
		$\done \leftarrow \Delta_{1}, \dtwo \leftarrow \Delta_{2}$ \;
		$\Delta_1 \leftarrow (1 - \epsilon) \Delta_1$ \;
		$\Delta_2 \leftarrow \sN \cdot \Delta_1 / |\tOUT|$ \;
		$\light \leftarrow T_I \cdot \sum(y_{\Delta_1}) + \cdot T_I \cdot \sum(x_{\Delta_2}) +$ \\ $ \hspace{3.3em} T_m \cdot |\domain(x)|  + T_s \cdot \countx(y_{\Delta_1}) \cdot |\domain(x)|$ \label{light:cost} \;
		$u,v,w \leftarrow \#\text{heavy }x,y,z \text{ values using } \count(w_\delta)$  \;
		$\heavy \leftarrow \hat{M}(u,v,w,co) + T_m \cdot (u \cdot v + u\cdot w)$ \;
		
		\If{$\prevl + \prevh \leq \light + \heavy$}{
			\KwRet{$\done, \dtwo$}
		}
		
	}
	\caption{Cost Based Optimizer}
	\label{algo:optimizer}
\end{algorithm}
\vspace{-0.1em}

To find the best possible estimates for $\Delta_1, \Delta_2$, we employ binary search over the value of $\Delta_2$. In each iteration, we increase or decrease its value by a factor of $(1 - \epsilon)$ where $\epsilon$ is a constant \footnote{We fix $\epsilon = 0.95$ for our experiments}. Once we fix the value of $\Delta_1$ and $\Delta_2$, we can query our precomputed index structure to find the exact number of operations that will be performed for all light $y$ values and all light $x$ values. Then, we find the number of heavy remaining values and get the estimate for time required to compute the matrix product. At the beginning of the next iteration, we compare the new time estimates with the previous iteration. If the new total time is larger than that of the previous iteration, we stop the process and use the last computed values as the degree thresholds. The entire process terminates in worst-case $O( \log^2 N)$ steps.

So far, we have not discussed how to estimate $\hat{M}(u,v,w,co)$. Since this quantity is system dependent, we precompute a table that stores the time required for different values of $u,v,w,co$. As a brute-force computation for all possible values is very expensive to store and compute, we store the time estimate for $\hat{M}(p,p,p,co)$ for $p \in \{1000, 2000, \dots, 20000\}, co \in [5]$. Then, given an arbitrary $u,v,w,co$, we can extrapolate from the nearest estimate available from the table. This  works well since \textsf{Eigen} implements the naive $O(n^3)$ (with optimizations) algorithm that offers predictable running time. \autoref{fig:mmul:scale} shows scalability of \textsf{Eigen} as the input matrix size increases. Since \textsf{Eigen} makes heavy use of SIMD instructions and vectorization, the running time displays a near quadratic growth rather than cubic for dimensions up to $5000 \times 5000$, beyond which the running time growth becomes cubic. 

\section{System Implementation} 
\label{sec:implementation}

We implement our techniques in \textsf{C++} as a standalone library. To perform matrix multiplication, we use \textsf{Eigen}~\cite{eigenweb} with \textsf{Intel MKL}~\cite{intelmkl} as the underlying linear algebra framework. We choose \textsf{Eigen} for its ease of use and its seamless support for parallelization, even though other frameworks such as \oldtt{MATLAB} are faster. \textsf{Intel MKL} offers two different functions for performing matrix operations: \oldtt{SGEMM} and \oldtt{DGEMM}. \oldtt{SGEMM} allows low precision real arithmetic while \oldtt{DGEMM} is for high precision arithmetic. This also makes \oldtt{DGEMM} $\mathbf{3x}$ slower than \oldtt{SGEMM} for the same operation being performed. We use floating point matrices everywhere rather than double precision or integer matrices for better performance.

At this point, we also wish to draw the attention of the reader towards some low level details of the \textsc{SSJ} and \textsf{SCJ} implementation. Since the goal is to output the result in arbitrary order, both implementations enumerate the result without storing any of the output. Enumerating the result in (say) decreasing order of similarity size or containment size will require storing the output and sorting it before providing the user with a pointer to the result. We implement this in the straightforward way by sorting \oldtt{std::vector} containing the output. Next, we describe the details of deduplication in our implementation for the case of unordered enumeration.

\smallskip
\noindent \introparagraph{Deduplication} Since matrix multiplication deduplicates the output for all heavy values, we only need to handle deduplication for the remaining output tuples. The straightforward way to deduplicate is to use a hashmap. However, this has two disadvantages: ${(i)}$ The memory for hashmap needs to be reserved upfront. This is critical to ensure that there is no resizing (and reshashing of the keys already present) of the hashmap at any point; ${(ii)}$ upfront reservation  would require $|\tOUT|$ amount of memory for deduplication, which is expensive both in terms of time and memory. 

\lstset{language=C++,
	basicstyle=\scriptsize	\ttfamily,
	keywordstyle=\color{blue}\ttfamily,
	stringstyle=\color{red}\ttfamily,
	commentstyle=\color{red}\ttfamily,
	morecomment=[l][\color{magenta}]{\#},
	autogobble=true,
	gobble=4,
	escapechar=|,
	numbers=left
}
\begin{lstlisting}
std::vector<int> y_light; // all light y values
std::unordered_map<int, set> R_xy; // indexed relation
std::unordered_map<int, set> R_yx; // indexed relation
std::vector<int> dedup(N); // reserving N memory |\label{line:assign}|
for(auto x : [N]) {
dedup.assign(N,0); 
for(auto y : y_light) {
if(R_xy[x].find(y) != R_xy[x].end()) {
for(auto z : R_yx[y]) {
dedup.at(z) += 1;
if (dedup.at(z) == 1) {
std::cout << x << z;
}
}
}
}
}
\end{lstlisting}

The code snippet above shows the join for all light $y$ values, which is the estimate used in~\autoref{light:cost} of~\autoref{algo:optimizer}. Line~\ref{line:assign} reuses the \texttt{dedup} vector to check that a given $z$ values has already been output or not. This is possible because we have fixed an $x$ value and then merge all the $z$ values reachable from $y$ that are connected to $x$. Since the above approach involves random access over the \texttt{dedup} vector, it can be easily an order of magnitude more expensive than serial access if the vector does not fit in the L1 cache. An alternative approach is to deduplicate by appending all reachable $z$ values, followed by sorting to deduplicate. For our experiments, we choose the best of the two strategies, depending on the number of elements that need to be deduplicated and the domain size of variables.

\begin{figure*}
	\captionsetup[subfigure]{oneside}
	\subcaptionbox{Two path query - single core \label{fig:join:twopath:singlecore}}%
	[0.30\linewidth]{\includegraphics[scale=0.6]{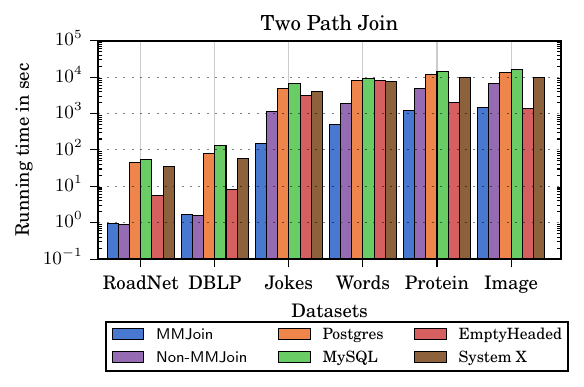}}
	\hfill	
	\subcaptionbox{Three star query - single core \label{fig:join:threestar:singlecore}}%
	[0.30\linewidth]{\includegraphics[scale=0.6]{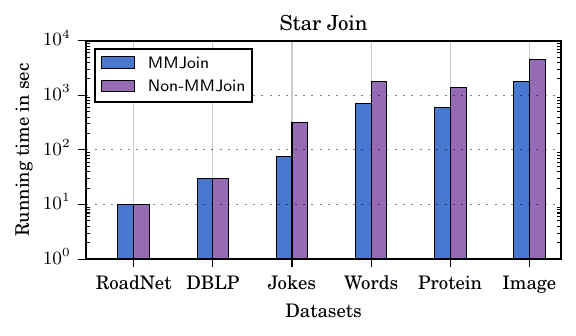}}
	\hfill
	\subcaptionbox{\textsf{SCJ} Running Time \label{fig:scj}}%
	[0.30\linewidth]{\includegraphics[scale=0.6]{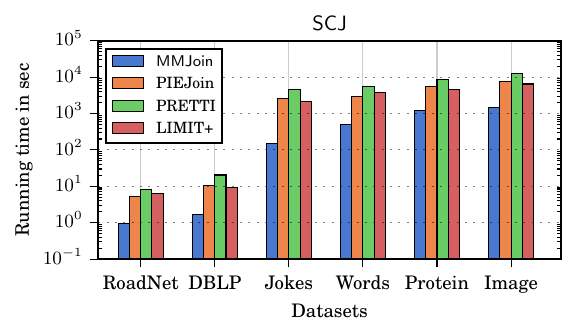}}
	\subcaptionbox{Jokes - multi core \label{fig:join:twopath:parallel:jokes}}%
	[0.22\linewidth]{\includegraphics[scale=0.6]{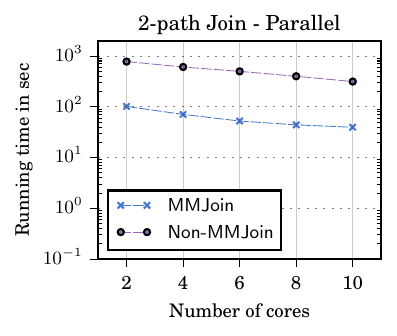}}
	\hfill		
	\subcaptionbox{Words - multi core \label{fig:join:twopath:parallel:words}}%
	[0.22\linewidth]{\includegraphics[scale=0.6]{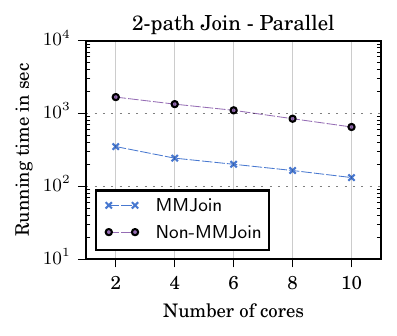}}
	\hfill	
	\subcaptionbox{Jokes - multi core \label{fig:join:threestar:parallel:jokes}}%
	[0.22\linewidth]{\includegraphics[scale=0.6]{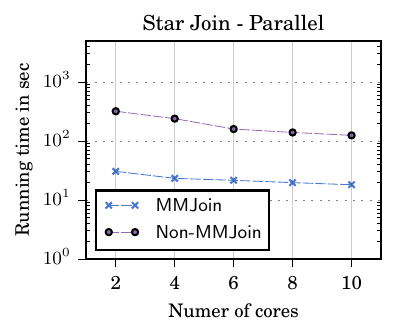}}
	\hfill		
	\subcaptionbox{Words - multi core \label{fig:join:threestar:parallel:words}}%
	[0.22\linewidth]{\includegraphics[scale=0.6]{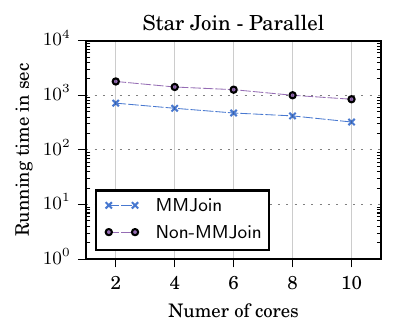}}
	\caption{Join Processing for two path and star query}	
\end{figure*} 

\smallskip
\noindent \introparagraph{Parallelization} The single-threaded execution of all algorithms easily reaches several hours when faced with gigabyte sized data sets and thus, parallel processing becomes necessary. \textsf{Eigen} parallelizes the matrix multiplication part in a coordination-free way, allowing both parts of our implementation to be highly parallelizable.  \autoref{fig:mmul:parallel} shows the running time of matrix multiplication as the number of cores increases. The speedup obtained is near linear as the resources available increases. This is possible because each core calculates the matrix product of a partition of data and requires no interaction with the other tasks.

\section{Experimental Evaluation} \label{sec:exp}

In this section, we empirically evaluate the performance of our algorithms. The main goal of the section is four fold: 

\begin{enumerate}
	\item Empirically verify the speed-up obtained for the 2-path and star queries using algorithm from~\autoref{sec:joinproject} compared to Postgres, MySQL, EmptyHeaded~\cite{aberger2017emptyheaded} and Commercial database X.
	\item Evaluate the performance of the two-path query against \texttt{SizeAware} and \texttt{SizeAware++} for unordered and ordered \textsf{SSJ}.
	\item Evaluate the performance of the 2-path query against three state-of-the-art algorithms, namely \textsf{PIEJoin} \cite{kunkel2016piejoin},  LIMIT+ \cite{bouros2016set}, PRETTI for \textsf{SCJ}.
	\item Validate the batching technique for boolean set intersection.
\end{enumerate}

All experiments are performed on a machine with Intel Xeon CPU E5-2660@2.6GHz, $20$ cores and $150$ GB RAM. Unless specified, all experiments are single threaded implementations. 
We use the open-source implementation of each algorithm. For all experiments, we focus on self-join i.e all relations are identical. All \textsf{C++} code is compiled using clang $8.0$ with \oldtt{-Ofast} flag and all matrix multiplication related code is additionally compiled with \oldtt{-mavx -mfma} \oldtt{-fopenmp} flags for multicore support. Each experiment is run $5$ times and we report the running time by averaging three values after excluding the slowest and the fastest runtime.

\begin{figure*}
	\captionsetup[subfigure]{oneside}
	\subcaptionbox{DBLP - single core \label{fig:ssj:dblp}}%
	[0.22\linewidth]{\includegraphics[scale=0.6]{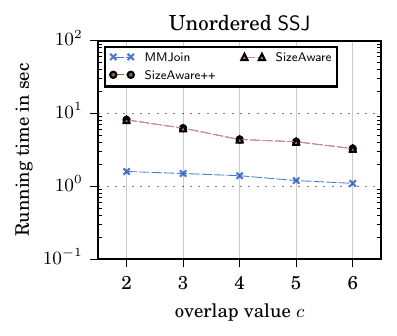}}
	\hfill
	\subcaptionbox{Jokes - single core \label{fig:ssj:jokes}}%
	[0.22\linewidth]{\includegraphics[scale=0.6]{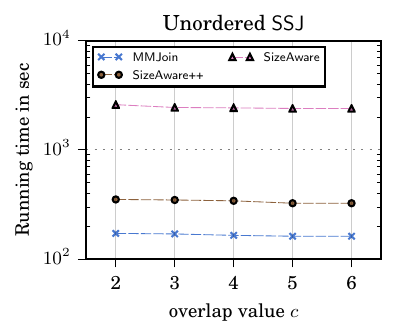}}
	\hfill
	\subcaptionbox{Image - single core \label{fig:ssj:image}}%
	[0.22\linewidth]{\includegraphics[scale=0.6]{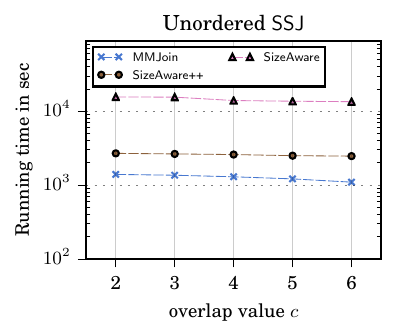}}
	\hfill
	\subcaptionbox{DBLP - multi core \label{fig:ssj:dblp:parallel}}%
	[0.22\linewidth]{\includegraphics[scale=0.6]{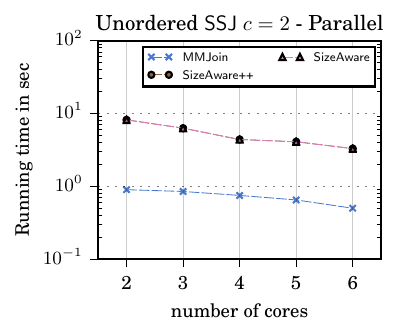}}
	
	\bigskip
	
	\subcaptionbox{DBLP - single core \label{fig:ssj:dblp:ordered}}%
	[0.22\linewidth]{\includegraphics[scale=0.6]{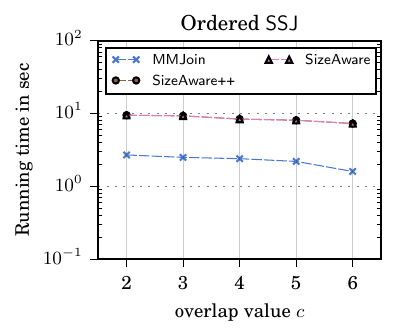}}
	\hfill
	\subcaptionbox{Jokes - single core \label{fig:ssj:jokes:ordered}}%
	[0.22\linewidth]{\includegraphics[scale=0.6]{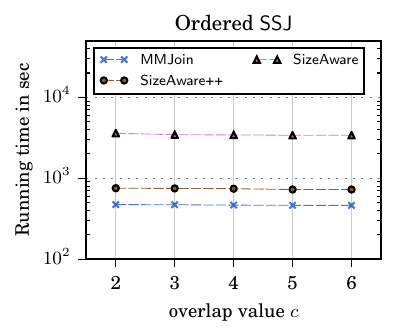}}
	\hfill
	\subcaptionbox{Jokes - multi core \label{fig:ssj:jokes:parallel}}
	[0.22\linewidth]{\includegraphics[scale=0.6]{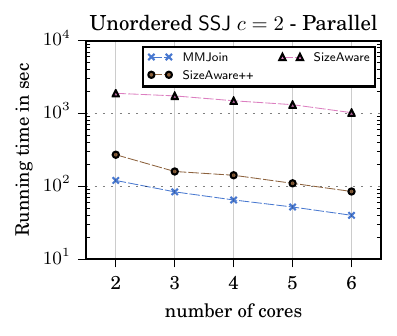}}
	\hfill
	\subcaptionbox{Image - multi core \label{fig:ssj:image:parallel}}%
	[0.22\linewidth]{\includegraphics[scale=0.6]{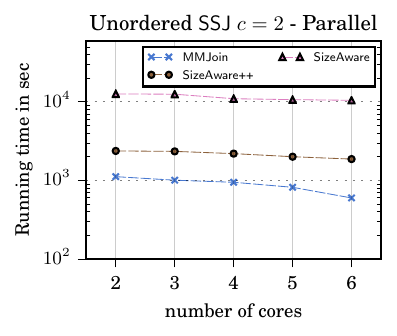}}
	\caption{Unordered and Ordered \textsf{SSJ}}	
\end{figure*} 

\subsection{Datasets}

We conduct experiments on six real-world datasets from different domains. DBLP~\cite{dblp} is a bibliography dataset from DBLP representing authors and papers. RoadNet~\cite{roadnet} is road network of Pennsylvania. Jokes~\cite{jokes} is a dataset scraped from Reddit where each set is a joke and there is an edge between joke and a word if the work is present in the joke. Words~\cite{words} is a bipartite graph between documents and the lexical tokens present in them. Image~\cite{image} dataset is a graph where each image is connected to a feature attribute if the image contains the corresponding attribute and Protein~\cite{protein} refers to a bipartite graph where an edge signifies interaction between two proteins. ~\autoref{table:dataset:characteristics} shows the main characteristics of the datasets. DBLP and RoadNet are examples of sparse datasets whereas the other four are dense datasets.

%

\begin{table}
	\scalebox{0.80}{
		\def\arraystretch{1.1}%
		\begin{small}
			\begin{tabular}{@{}lrrrccc@{}}\toprule
				Dataset & $|R|$ & No. of sets & $|\domain|$ & Avg set size & Min set size & Max set size \\ \midrule
				\textbf{DBLP} &  10M & 1.5M  & 3M & 6.6 & 1 & 500  \\ \hline
				\textbf{RoadNet} &  1.5M & 1M & 1M & 1.5 & 1 & 20   \\ \hline
				\textbf{Jokes} & 400M & 70K & 50K & 5.7K & 130 & 10K \\ \hline
				\textbf{Words} & 500M & 1M & 150K & 500 & 1 & 10K \\ \hline 
				\textbf{Protein} & 900M & 60K & 60K & 15K & 50 & 50K  \\ \hline
				\textbf{Image} & 800M & 70K & 50K  & 11.4K & 10K & 50K \\ \hline
				\bottomrule
			\end{tabular}
	\end{small}}
	\caption{Dataset Characteristics}
	\label{table:dataset:characteristics}
\end{table}

\subsection{Simple Join Processing}

In this part, we evaluate the running time for the two queries: $\mtwopath$ and $Q^\star_3$. To extract the maximum performance from Postgres, we use PGTune to set the best configuration parameters. This is important to ensure that the query plan does not perform nested loop inner joins. For all datasets, we create a hash index over each variable to ensure that the optimizer can choose the best query plan. We manually verify that query plan generated by PostgreSQL (and MySQL) when running these queries chooses HashJoin or MergeJoin. \hlrtwo{For X, we allow  up to 1TB of disk space and supply query hints to make sure that all of the CPU, RAM memory is available for query execution}.

~\autoref{fig:join:twopath:singlecore} shows the run time for different algorithms on a single core. MySQL and Postgres have the slowest running time since they evaluate the full query join result and then deduplicate. \hlrtwo{DBMS X performs marginally better than MySQL and Postgres.} Non-matrix multiplication (denoted \textsf{Non-MMJoin}) join  based on~\autoref{lem:star:basic} is the second best algorithm. Matrix multiplication based join (denoted \textsf{MMJoin}) is the fastest on all datasets except RoadNet and DBLP, where the optimizer chooses to compute the full join. \hlrone{A key reason for the huge performance difference between \textsf{MMJoin} and other algorithms is that deduplication by computing the full join result requires either sorting the data or using hash tables, both of which are expensive operations. In particular, using hash tables requires rehashing of entires every time the hash table increases. Similarly, sorting the full join result is expensive since the full join result can be orders of magnitude larger than the projection query result. Matrix multiplication avoids this since worst-case optimal joins can efficiently process the light part of the input and matrix multiplication is space efficient due to its implicit factorization of the output formed by heavy values. Remarkably, EmptyHeaded performs comparable to \textsf{MMJoin} for Jokes dataset and outperforms \textsf{MMJoin} slightly on Image dataset. This is because Image dataset exclusively contains a dense component where every the output is close to a clique. Since EmptyHeaded is designed as a linear algebra engine like Intel MKL, the performance is very similar.}
~\autoref{fig:join:twopath:parallel:jokes} and~\ref{fig:join:twopath:parallel:words} show the performance of the combinatorial and non-combinatorial algorithm as the number of cores increases. Both algorithms show a speed-up. We omit MySQL and Postgres since they do not allow for multicore processing of single queries. 

Next, we turn to the star query on three relations. For this experiment, we take the largest sample of each relation so that the result can fit in main memory and the join finishes in reasonable time. ~\autoref{fig:join:threestar:singlecore} shows the performance of the combinatorial and non-combinatorial join on a single core. All other engines (except EmptyHeaded) failed to finish in $15000$ seconds except on RoadNet and DBLP. EmptyHeaded performed similarly to \textsf{MMJoin} on Protein and Image datasets but not on other datasets.~\autoref{fig:join:threestar:parallel:jokes} and~\ref{fig:join:threestar:parallel:words} show the performance in a multicore setting for Jokes and Words datasets.  Once again, matrix multiplication performs better than its combinatorial version across all experiments.

\subsection{Set Similarity}

In this section, we look at set similarity (SSJ).  For both settings below, we materialize the output at all nodes in the prefix tree. We will compare the performance of \textsf{MMJoin}, \textsf{SizeAware} and \textsf{SizeAware++}. We begin with the unordered setting.

\smallskip
\noindent \introparagraph{Unordered \textsf{SSJ}} ~\autoref{fig:ssj:dblp},~\ref{fig:ssj:jokes} and~\ref{fig:ssj:image} show the running time of \textsf{MMJoin}, \textsf{SizeAware} and \textsf{SizeAware++} on a single core for DBLP, Jokes and Image dataset respectively. Since DBLP is a sparse dataset with small set sizes, \textsf{MMJoin} is the fastest and both \textsf{SizeAware} and \textsf{SizeAware++} are marginally slower due to the optimizer cost. For Jokes and Image datasets, \textsf{SizeAware} is the slowest algorithm. This is because both the light and heavy processing have a lot of deduplication to perform. \textsf{SizeAware++} is an order of magnitude faster than \textsf{SizeAware} since it uses matrix multiplication but is slower than \textsf{MMJoin} because it still needs to enumerate the $c$-subsets before using matrix multiplication. \textsf{MMJoin} is the fastest as it is output sensitive and performs the best in a setting with many duplicates. Next, we look at the parallel version of unordered \textsf{SSJ}. ~\autoref{fig:ssj:dblp:parallel},~\ref{fig:ssj:jokes:parallel} and~\ref{fig:ssj:image:parallel} show the results for multi core settings. For each experiment, we fix the overlap to $c=2$. Observe that \textsf{MMJoin} join and \textsf{SizeAware++} are more scalable than \textsf{SizeAware}. This is because the light sets processing of \textsf{SizeAware} cannot be done in parallel while matrix multiplication based deduplication can be performed in parallel.
\begin{figure*}
	\captionsetup[subfigure]{oneside}
	\subcaptionbox{Image - single core \label{fig:ssj:image:ordered}}%
	[0.22\linewidth]{\includegraphics[scale=0.6]{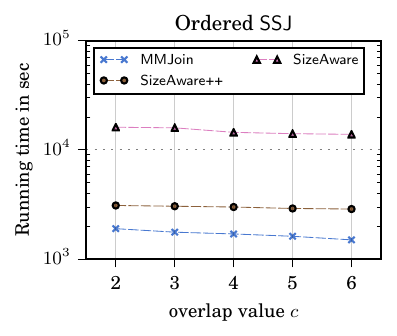}}
	\hfill
	\captionsetup[subfigure]{oneside}
	\subcaptionbox{Jokes \label{fig:delay:jokes}}%
	[0.22\linewidth]{\includegraphics[scale=0.6]{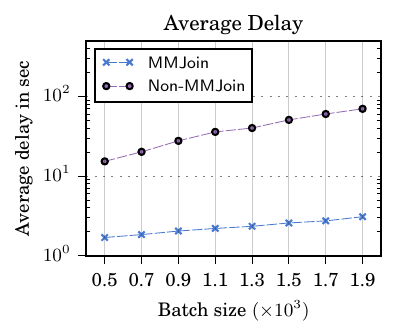}}
	\hfill
	\subcaptionbox{Words \label{fig:delay:words}}%
	[0.22\linewidth]{\includegraphics[scale=0.6]{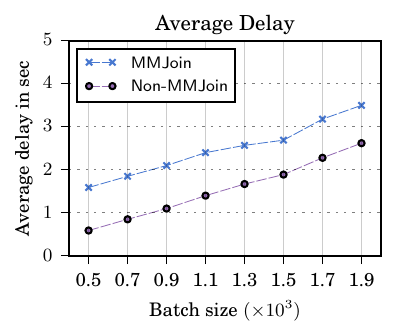}}
	\hfill
	\subcaptionbox{Image \label{fig:delay:image}}%
	[0.22\linewidth]{\includegraphics[scale=0.6]{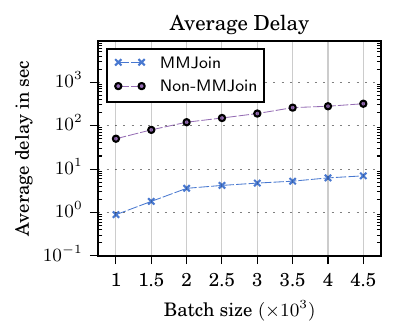}}
	\caption{Ordered \textsf{SSJ} and minimizing average delay}	
\end{figure*} 
\begin{figure*}[!htpb]
	\captionsetup[subfigure]{oneside}
	\subcaptionbox{Jokes dataset  \label{fig:scj:jokes:parallel}}%
	[0.22\linewidth]{\includegraphics[scale=0.6]{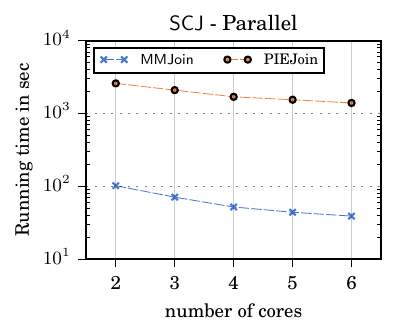}}
	\hfill
	\subcaptionbox{Words - multi core  \label{fig:scj:words:parallel}}%
	[0.22\linewidth]{\includegraphics[scale=0.6]{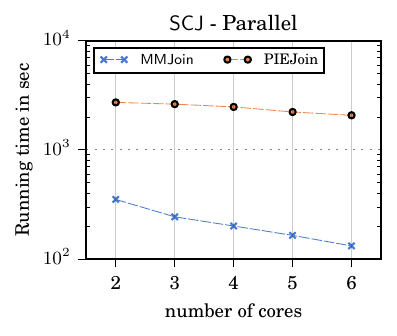}}
	\hfill
	\subcaptionbox{Protein - multi core \label{fig:scj:protein:parallel}}%
	[0.22\linewidth]{\includegraphics[scale=0.6]{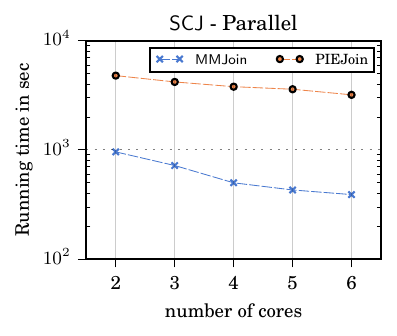}}
	\hfill
	\subcaptionbox{Image - multi core \label{fig:scj:image:parallel}}%
	[0.22\linewidth]{\includegraphics[scale=0.6]{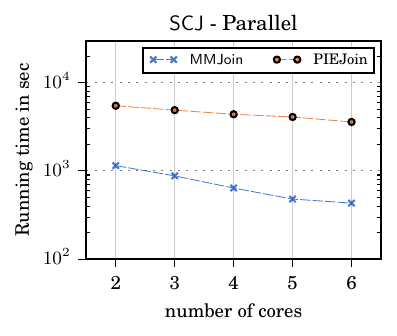}}
	\caption{Unordered Parallel \textsf{SCJ}}	
\end{figure*} 

\smallskip
\noindent \introparagraph{Ordered \textsf{SSJ}} Recall that for ordered \textsf{SSJ}, our goal is to enumerate the set pairs in descending order of set similarity. Thus, once the set pairs and their overlap is known, we need to sort the result using overlap as the key. ~\autoref{fig:ssj:dblp:ordered} and~\ref{fig:ssj:jokes:ordered} show the running time for single threaded implementation of ordered set similarity. Compared to the unordered setting, the extra overhead of materializing the output and sorting the result increases the running time for all algorithms. For \textsf{SizeAware}, there is an additional overhead of finding the overlap for all light sets as well. Both \textsf{MMJoin} and \textsf{SizeAware++} maintain their advantage similar to the unordered setting.

\begin{figure}[!htp]
	\includegraphics[scale=0.7]{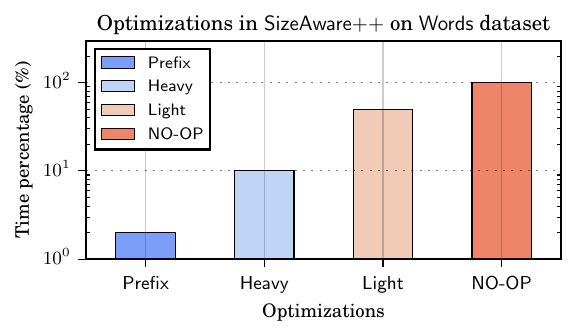}%
	\caption{SSJ - Impact of optimizations on Words}	\label{fig:optimizations}
\end{figure}

\smallskip
\noindent \introparagraph{Impact of optimizations} Recall that \textsf{SizeAware++} contains three main optimizations - processing heavy sets using \textsf{MMJoin}, processing light sets via \textsf{MMJoin} and using prefix based materialization for computation sharing.~\autoref{fig:optimizations} shows the effect of switching on various optimizations. \textsf{NO-OP} denotes all optimizations switched off. The running time is shown as a percentage of the \textsf{NO-OP} running time ($100\%$). \textsf{Light} denotes using two-path join on only light sets identified by \textsf{SizeAware} but not using the prefix optimizations. \textsf{Heavy} includes the \textsf{Light} optimizations switched on plus two-path join processing on the heavy sets but prefix based optimization is still switched off. Finally, \textsf{Prefix} switches on materialization of the output in prefix tree on top of \textsf{Light} and \textsf{Heavy}. As the figure shows, both \textsf{Light} and \textsf{Heavy} optimizations together improve the running time by an order of magnitude and \textsf{Prefix} further improves by a factor of $5 \times$.

\subsection{Set Containment}

In this section, we evaluate the performance of different set containment join algorithms.~\autoref{fig:scj} shows the running time of \textsf{PIEJoin}, PRETTI and LIMIT+. For all \textsf{SCJ} algorithms, we use the infrequent sort order, choose a limit value of two for LIMIT+ and run the variant where the output is materialized (instead of just simply counting its size). Once again join processing yields the fastest running time since the join output is a superset of the set containment join result and except RoadNet and DBLP, the join-project result and \textsf{SCJ} result is close to each other. Since the average set size is large for most datasets, \textsf{SCJ} algorithms need to perform expensive verification operations. For the parallel setting, ~\autoref{fig:scj:jokes:parallel},~\ref{fig:scj:words:parallel},~\ref{fig:scj:protein:parallel} and~\ref{fig:scj:image:parallel} show the performance of \textsf{PIEJoin} vs. \textsf{MMJoin}. \textsf{PIEJoin} does not scale as well as \textsf{MMJoin} as it is sensitive to data distribution and choice of partitions chosen by the heuristic in the algorithm.

\subsection{Boolean Set Intersection}

In this part of the experiment, we look at the boolean set intersection scenario where queries are arriving in an online fashion.  The arrival rate of queries is set to $B=1000$ queries per second and our goal is to minimize the average delay metric as defined in Section~\ref{sec:bsi}. The workload is generated by sampling each set pair uniformly at random. We run this experiment for $300$ seconds for each batch size and report the mean average delay metric value. ~\autoref{fig:delay:jokes},~\ref{fig:delay:words} and~\ref{fig:delay:image} show the average delay for the three datasets at different batch sizes. Recall that the smaller batch size we choose, the more number of processing units are required. For the Jokes dataset, \textsf{Non-MMJoin} has the smallest average delay of $\approx 1s$ when $S = 10$. In that time, we collect a further $1000$ requests, which means that there is a need for $100$ parallel processing units. On the other hand, \textsf{MMJoin} achieves a delay of $\approx 2s$ at batch size $900$. Thus, we need only $3$ parallel processing units in total to keep up with the workload while sacrificing only a small penalty in latency. For the Image dataset, \textsf{MMJoin} can achieve average delay of $1s$ at $S = 1000$ queries while \textsf{Non-MMJoin} achieves $50s$ at the same batch size. This shows that matrix multiplication is useful for achieving a smaller latency using less resources, in line with the theoretical prediction. For the Words dataset, most of the sets have a small degree. Thus, the optimizer chooses to evaluate the join via the combinatorial algorithm. This explains the in sync behavior of average delay for both the algorithms. Note that \textsf{MMJoin} is marginally slower because of the overhead of the optimizer ($\leq 2s$).

\section{Related Work} \label{sec:related}

Theoretically,~\cite{DBLP:conf/csl/BaganDG07} and~\cite{amossen2009faster} are the most closely related works to our considered setting (as discussed in~\autoref{sec:known}). In practice, most of the previous work has considered join-project query evaluation by pushing down the projection operator in the query plan~\cite{gupta1995generalized, bhargava1997enumerating, gupta1995aggregate, ceri1985translating}. \hlrtwo{LevelHeaded~\cite{aberger2018levelheaded} and EmptyHeaded~\cite{aberger2017emptyheaded} are general linear algebra systems that use highly optimzed set intersections to speed up evaluation of cyclic joins, counting queries and support projections over them. Since Intel \textsf{MKL} is also a linear algebra library, one can also use EmptyHeaded as the underlying framework for performing matrix multiplication.}
Very recently,~\cite{kara2019trade} made significant progress by providing algorithms that tradeoff pre-processing time and worst-case delay guarantees for heirarchical queries (star join is a subset of heirarchical queries). The main result states that for star query with $k$ relations, there exists an algorithm that pre-processes in time $T = O(N^{1 + (k-1) \epsilon})$ such that it is possible to enumerate the join-project result without duplications with worst-case delay guarantee $ \delta = O(N^{1 + \epsilon})$ for any $\epsilon \in [0,1]$. This implies that the total running time is bounded by $\delta \cdot |\tOUT|$.
For group-by aggregate queries,~\cite{xirogiannopoulos2019memory} also used worst-case optimal join algorithms to avoid evaluating binary joins at a time and materializing the intermediate results. However, the running time of their algorithm is not output sensitive with respect to the final projected result and could potentially be improved upon by using our proposed ideas.

\section{Conclusion and Future Work} \label{sec:conclusion}

In the paper, we study the evaluation of join queries with projections. This is useful for a wide variety of tasks including set similarity, set containment and boolean query answering. We describe an algorithm based on fast matrix multiplication that allows for theoretical speedups. Empirically, we demonstrated that the framework is also practically useful and can provide speedups of up to $50 \times$ for some datasets. There are several promising future directions that remain to be explored. The first key direction is to extend our techniques to arbitrary acyclic queries with projections. In order to do so, we need better join-project size estimation techniques and building a query plan that decomposes the join into multiple subqueries and evaluates in the optimal way. This can potentially be done by modifying estimators for set union and set intersection such as KMV and HyperLogLog. Second, it remains unclear if the same techniques can also benefit cyclic queries or not. For instance, AYZ algorithm is applicable to counting cycles in graph using matrix multiplication. It would be interesting to extend the algorithm to enumerate join-project output where the user can choose arbitrary projection variables on an cyclic query. It will also be interesting to see if fast matrix multiplication can help in group-by aggregate queries for longer path queries.
	
	\bibliographystyle{abbrv}
	\bibliography{reference}

\end{document}